\pdfoutput=1
\documentclass[a4paper,11pt]{article}
\usepackage{jheppub,color,ulem,xspace}

\def\bpm{\begin{pmatrix}} 
\def\epm{\end{pmatrix}} 
\def\bea{\begin{eqnarray}}
\def\eea{\end{eqnarray}}
\def\nn{\nonumber}
\def\be{\begin{equation}}
\def\ee{\end{equation}}
\def\bsp#1\esp{\begin{split}#1\end{split}} 

\definecolor{mkgreen}{rgb}{0.2,.70,.3}

\newcommand{\gR}{g_{R}}
\newcommand{\hppmm}{$H^{\pm \pm}$\xspace}
\newcommand{\lamR}{\lambda_R}
\newcommand{\lamS}{\lambda_S}
\newcommand{\mhpp}{m_{H^{\pm \pm}}}

\newcommand{\veinsr}{v_{1R}}
\newcommand{\vzweir}{v_{2R}}
\newcommand{\vu}{v_u}
\newcommand{\vd}{v_d}
\newcommand{\vs}{v_S}

\newcommand{\spheno}{\texttt{SPheno}\xspace}
\newcommand{\sarah}{\texttt{SARAH}\xspace}
\newcommand{\vevacious}{\texttt{Vevacious}\xspace}

\newcommand{\madgraph}{\texttt{MadGraph5\_aMC@NLO}\xspace}
\newcommand{\feynrules}{\texttt{FeynRules}\xspace}
\newcommand{\ssp}{\texttt{SSP}\xspace}

\newcommand{\eq}[1]{eq.~\eqref{#1}}
\newcommand{\fig}[1]{figure~\ref{#1}}

\newcommand{\figs}[1]{figures~\ref{#1}}
\newcommand{\Figs}[1]{Figures~\ref{#1}}
\newcommand{\REF}[1]{ref.~\cite{#1}}
\newcommand{\REFS}[1]{refs.~\cite{#1}}
\newcommand{\SEC}[1]{section~\ref{#1}}

\newcommand{\oneloop}{one-loop\xspace}

\def\lsim{\raise0.3ex\hbox{$\;<$\kern-0.75em\raise-1.1ex\hbox{$\sim\;$}}}
\def\gsim{\raise0.3ex\hbox{$\;>$\kern-0.75em\raise-1.1ex\hbox{$\sim\;$}}}

\newcommand{\AddrStrasbourg}{  Institut Pluridisciplinaire Hubert Curien/D\'epartement
    Recherches Subatomiques,\\ Universit\'e de Strasbourg/CNRS-IN2P3,
    23 rue du Loess, F-67037 Strasbourg, France}

\newcommand{\AddrWue}{%
Institut f\"ur Theoretische Physik und Astronomie, 
Universit\"at W\"urzburg\\
Emil-Hilb-Weg 22, 97074 W\"urzburg, Germany}

\allowdisplaybreaks

\title{Doubly-charged Higgs and vacuum stability in left-right supersymmetry}

\author[a]{Lorenzo Basso}
\author[a]{, Benjamin Fuks}
\author[b]{, Manuel E. Krauss}
\author[b]{and Werner Porod} 

\affiliation[a]{\AddrStrasbourg}
\affiliation[b]{\AddrWue}

\emailAdd{lorenzo.basso@iphc.cnrs.fr}
\emailAdd{benjamin.fuks@iphc.cnrs.fr}
\emailAdd{manuel.krauss@physik.uni-wuerzburg.de}
\emailAdd{porod@physik.uni-wuerzburg.de}

\abstract{
We present an analysis of supersymmetric left-right symmetric models with Higgs
fields lying in the adjoint representation of $SU(2)_R$. These models feature
a doubly-charged Higgs boson which gets its mass only at the loop level. We
present, for the first time, a complete \oneloop calculation of this mass and
show that contributions that have been neglected so far can shift it
by a few hundreds of GeV. We then combine this observation with LHC
bounds deduced from extra charged gauge boson and doubly-charged Higgs boson
searches. In particular, we point out that existing limits get substantially
modified by the presence of singly-charged Higgs bosons that are also predicted
in these models. In addition, we address constraints stemming from vacuum
stability considerations and show how the considered class of models could be
ruled out at the next LHC run in the absence of any signal.
}


\begin{document}
\maketitle


\section{Introduction}
The Large Hadron Collider (LHC) at CERN has been built to explore the TeV energy
scale and pin down the mechanism responsible for the breaking of the electroweak
symmetry. With this respect, the discovery of a Higgs boson consistent with the
Standard Model of
particle physics \cite{ATLAS:2012ae,Chatrchyan:2012tx} can be seen as a first
success of the LHC physics programme. This also consists of the
first observation of a particle intrinsically unstable with respect to quantum
corrections. One can therefore expect either an unnatural fine-tuning
or a stabilization arising from physics beyond the Standard Model that would
emerge at a scale reachable at present and future colliders. Although we do not
know which theory could be the most suitable, a lot
of efforts have been put, during the last decades, in the study of a plethora of
new physics models. The Minimal Supersymmetric Standard Model~\cite{
Nilles:1983ge,Haber:1984rc} (MSSM) is one of the most popular of these and it
provides, additionally to curing the infamous hierarchy problem, a solution for
other conceptual issues plaguing the Standard Model.

There is currently no evidence for supersymmetry and limits on the superpartners
of the Standard Model particles are hence pushed to higher and higher scales.
Most results have however been derived either in the MSSM or within simplified
models inspired by the latter. There are nevertheless large varieties of
alternative supersymmetric realizations that could evade all current bounds and
deserve to be investigated. For instance, the MSSM inherits some of the flaws of
the Standard Model and extending the model gauge symmetry could provide a
mechanism yielding the generation of the neutrino masses and an explanation for
the strong $CP$ problem.

We consider in this work supersymmetric theories that exhibit a left-right
symmetry~\cite{Francis:1990pi,Huitu:1993uv,Huitu:1993gf,Babu:2008ep} and that
are based on an $SU(3)_c \times SU(2)_L \times SU(2)_R \times
U(1)_{B-L}$ gauge group. They provide a solution for the hierarchy problem,
explain the smallness of the neutrino masses, can include a dark matter
candidate, and they potentially solve the strong
$CP$ problem because of parity invariance at the fundamental level.
Moreover, minimal supersymmetric setups like constrained versions of the MSSM
cannot usually predict, at the tree level, a neutral Higgs boson mass compatible
with the observations and require large higher-order contributions.
This issue can be alleviated by extending the Higgs sector, like
in left-right supersymmetric models where the spontaneous breaking of the
left-right symmetry all the way down to the $U(1)$ electromagnetic group
necessitates several Higgs fields~\cite{Babu:1987kp,Zhang:2008jm,
Krauss:2013jva}.

Several left-right supersymmetric setups have been studied in the past, with
different Higgs sectors and symmetry breaking details. We choose to focus on a
class of minimal models featuring two $SU(2)_R$ Higgs triplets that are
responsible for the breaking of $SU(2)_R\times U(1)_{B-L}$ to the electroweak
group, two extra $SU(2)_L$ Higgs triplets that allow for parity invariance at
higher scales and two $SU(2)_L\times SU(2)_R$ bidoublets yielding both the
breaking of the electroweak symmetry down to electromagnetism and the generation
of acceptable fermion masses and mixings~\cite{Kuchimanchi:1993jg,
Kuchimanchi:1995vk,Chacko:1997cm}. In addition, we include an extra singlet
field to achieve a successful electroweak symmetry breaking in the
supersymmetric limit~\cite{Babu:2008ep}. With such a choice for the Higgs
sector, the interactions of the lepton supermultiplets with the triplets are
sufficient for generating neutrino masses through a see-saw
mechanism~\cite{Mohapatra:1979ia}, and the presence of the singlet
solves the so-called $\mu$-problem of the MSSM as all bilinear terms of the
superpotential can be generated dynamically. These models are currently probed
at the LHC through traditional supersymmetry~\cite{atlassusy,cmssusy} and extra
gauge boson searches~\cite{atlasexo,cmsexo}, as well as analyses dedicated to
the doubly-charged (Higgs and Higgsino) states induced by the $SU(2)_L$ and
$SU(2)_R$ triplets~\cite{atlasexo,cmsexo}. The observation of such
doubly-charged particles would indeed provide valuable information about the
symmetry breaking pattern of the underlying theory.

The minimization of the scalar potential of the model is known to lead to a
tree-level solution in which either $R$-parity invariance~\cite{
Kuchimanchi:1993jg,Kuchimanchi:1995vk} or electric charge conservation~\cite{
Chacko:1997cm} is lost. Including
\oneloop heavy Majorana neutrino contributions however allows for
satisfactory solutions to the minimization equations~\cite{Babu:2008ep,
Babu:2014vba}. In this case, the right-handed sneutrino fields are prevented
from getting non-vanishing vacuum expectation values so that $R$-parity is
conserved and no dangerous lepton number violating operators appear in the
superpotential. Additionally, these radiative corrections also protect the
doubly-charged Higgs bosons from being tachyonic and breaking
electromagnetism. This has been demonstrated both in pioneering calculations
relying on the Coleman-Weinberg effective potential~\cite{Babu:2008ep} and in a
more recent calculation using a diagrammatic approach in the gaugeless limit
where all contributions proportional to the gauge couplings are
neglected~\cite{Babu:2014vba}.

In this work, we start with a brief description of the considered left-right
supersymmetric models (\SEC{sec:model}) and then show,
in \SEC{sec:hpp}, that the previously calculated corrections to the
doubly-charged Higgs mass are not sufficient to render it compatible with the
current experimental bounds~\cite{Chatrchyan:2012ya}. We subsequently evaluate
the complete \oneloop result and demonstrate that the doubly-charged
Higgs mass can be increased well above the present limits. Other recent LHC
constraints are investigated in \SEC{sec:LHCbounds} and we
further study the minimum of the scalar potential at the \oneloop level in
\SEC{sec:vacuum_stability}, analyzing the requirements for ensuring this
minimum to be global. Our conclusions about the phenomenological viability of
the model are given in \SEC{sec:conclusions}.


 \section{The Model}
\label{sec:model}
Many left-right supersymmetric setups have been developed in the past. We
adopt the configuration introduced in \REF{Babu:2008ep}. The
Standard Model left- and right-handed quarks (leptons) are embedded, together
with their scalar superpartners, in $Q_L$ and $Q_R$ ($L_L$ and $L_R$)
supermultiplets lying in the fundamental representation of the $SU(2)_L$ and
$SU(2)_R$ gauge groups, respectively. The Higgs(ino) sector is constituted of
two $SU(2)_L$ triplets ($\Delta_{1L}$ and $\Delta_{2L}$ with a $B\!-\!L$ charge
$Q_{B\!-\!L}=
\mp 2$), two $SU(2)_R$ triplets ($\Delta_{1R}$ and $\Delta_{2R}$ with
$Q_{B\!-\!L}=\mp 2$), two $SU(2)_L\times SU(2)_R$ bidoublets ($\Phi_1$ and
$\Phi_2$ not charged under the $B\!-\!L$ symmetry) and one gauge singlet ($S$).
As demonstrated in
pioneering works, this configuration allows for a successful symmetry breaking
of the $SU(2)_L\times SU(2)_R \times U(1)_{B-L}$ gauge symmetry down to the
$U(1)$ electromagnetic gauge group, it includes a see-saw mechanism
yielding the generation of the neutrino masses and it provides acceptable
fermion masses and mixings with respect to data~\cite{Mohapatra:1979ia,
Kuchimanchi:1993jg,Kuchimanchi:1995vk,Chacko:1997cm}.

Assuming a discrete $\mathbb{Z}_3$ symmetry, the superpotential only
contains terms that are trilinear in the fields,
\be\bsp
   & W =
   Q_L y^Q_1 \Phi_1 Q_R +
   Q_L y^Q_2 \Phi_2 Q_R +
   L_L y^L_1 \Phi_1 L_R
   + L_L y^L_2 \Phi_2 L_R +
   L_L y^L_3 \Delta_{2L} L_L
\\  &\
   + L_R y^L_4 \Delta_{1R} L_R 
   + \lambda_L S \Delta_{1L} \!\cdot\! \Delta_{2L}
   + \lamR     S \Delta_{1R} \!\cdot\! \Delta_{2R}
   + \lambda_1 S \Phi_1 \!\cdot\! \Phi_1
   + \lambda_2 S \Phi_2 \!\cdot\! \Phi_2
\\ &\
   + \lambda_{12} S \Phi_1 \!\cdot\! \Phi_2
   + \frac13 \lamS S^3\,,
\label{eq:Wtrip}
\esp\ee
where we refer to \REF{Alloul:2013fra} for details on
the underlying (understood) index structure. In those notations, the various
coupling strengths have been embedded into $3\times 3$ Yukawa matrices ($y^i_Q$
and $y^j_L$ with $i=1,2$ and $j=1,2,3,4$) and $\lambda$ parameters dictating the
size
of the Higgs self-interactions. The supersymmetry-breaking Lagrangian contains
sfermion and gaugino mass terms as well as trilinear ($T$) interactions whose
structure is given by their superpotential counterparts.

The spontaneous symmetry-breaking mechanism proceeds in two steps. The $SU(2)_R
\times U(1)_{B-L}$ group is first broken to the hypercharge group and the two
$SU(2)_R$ Higgs triplets develop non-vanishing vacuum expectation values (vevs).
In a second
step, the electroweak symmetry is broken to electromagnetism and the
bidoublet Higgs fields get non-zero vacuum expectation values. The vacuum
state is characterized by the vevs of all the neutral
components of the Higgs fields,
\bea
  \langle \Delta_{1L} \rangle =
  \left\langle \begin{pmatrix}
  \frac{\Delta_{1L}^-}{\sqrt{2}} & \Delta_{1L}^0 \\
  \Delta_{1L}^{--} & -\frac{\Delta_{1L}^-}{\sqrt{2}}
  \end{pmatrix}\right\rangle = 
  \begin{pmatrix}
  0 & \frac{v_{1L}}{\sqrt{2}} \\
  0 & 0
  \end{pmatrix} \ , \ \ &
  \langle \Delta_{2L} \rangle =
  \left\langle \begin{pmatrix}
  \frac{\Delta_{2L}^+}{\sqrt{2}} & \Delta_{2L}^{++} \\
  \Delta_{2L}^{0} & -\frac{\Delta_{2L}^+}{\sqrt{2}}
  \end{pmatrix}\right\rangle = 
  \begin{pmatrix}
  0 & 0\\
  \frac{v_{2L}}{\sqrt{2}}  & 0
  \end{pmatrix} \ , \nn \\ 
  \langle \Delta_{1R} \rangle =
  \left\langle \begin{pmatrix}
  \frac{\Delta_{1R}^-}{\sqrt{2}} & \Delta_{1R}^0 \\
  \Delta_{1R}^{--} & -\frac{\Delta_{1R}^-}{\sqrt{2}}
  \end{pmatrix}\right\rangle = 
  \begin{pmatrix}
  0 & \frac{v_{1R}}{\sqrt{2}} \\
  0 & 0
  \end{pmatrix} \ , \ \  &
  \langle \Delta_{2R} \rangle =
  \left\langle \begin{pmatrix}
  \frac{\Delta_{2R}^+}{\sqrt{2}} & \Delta_{2R}^{++} \\
  \Delta_{2R}^{0} & -\frac{\Delta_{2R}^+}{\sqrt{2}}
  \end{pmatrix}\right\rangle = 
  \begin{pmatrix}
  0 & 0\\
  \frac{v_{2R}}{\sqrt{2}}  & 0
  \end{pmatrix} \ , \nn \\
  \langle \Phi_1 \rangle =\Big\langle
  \begin{pmatrix}
  \Phi_1^0 & \Phi_1^+ \\
  \Phi_1^- & \Phi'{}_1^{0}
  \end{pmatrix} \Big\rangle = 
  \begin{pmatrix}
  \frac{v_d}{\sqrt{2}} & 0 \\
  0 & 0
  \end{pmatrix} \ ,& \ \
  \langle \Phi_2 \rangle =\Big\langle
  \begin{pmatrix}
  \Phi'{}_2^0 & \Phi_2^+ \\
  \Phi_2^- & \Phi_2^{0}
  \end{pmatrix} \Big\rangle = 
  \begin{pmatrix}
  0 & 0 \\
  0 & \frac{v_u}{\sqrt{2}}
  \end{pmatrix}\ ,\nn \\
   \langle S \rangle = \frac{v_S}{\sqrt{2}}\ , &
  \label{eq:desired_vacuum}
\eea
where all vevs are taken real and positive (as allowed by suitable field
redefinitions). Although the $\Phi'^0$ fields could
also develop non-zero vacuum expectation values giving rise to $W/W_R$ mixing,
these are strongly
constrained by kaon data so that they are ignored. Moreover, the phase of
$v_S$ cannot in principle be rotated away but we omit it for simplicity.
The remaining vevs can be further constrained.
Both the $v_{iR}$ vevs are imposed to be large by the masses
of the $W_R$ and $Z_R$ gauge bosons.
Furthermore, the smallness of the Standard Model neutrino masses and electroweak
precision data require that the vevs developed by the $SU(2)_L$ Higgs triplets
$v_{iL}$ are negligible\footnote{Another acceptable choice implies that
the $y_3^L$ couplings are tiny and the $y_4^L$ ones are large.
This is however very unnatural with the left-right symmetry.}.
Finally, the size of $v_S$ is related to
the effective $\mu$-terms of the superpotential, and
we consequently impose the hierarchy \mbox{$v_S, v_{1R}, v_{2R} \gg v_{u,d} \gg
v_{1L}\!\approx\! v_{2L} \!\approx\! 0$}. For further references,
we define \mbox{$v_{1R}^2 + v_{2R}^2 = v_R^2$},
\mbox{$\tan \beta_R = v_{2R}/v_{1R}$},
\mbox{$v_d^2 + v_u^2 = v^2$} and \mbox{$\tan \beta = v_u/v_d$} and we
additionally assume $\lambda_{1} = \lambda_2 = 0$, analogously to
\REFS{Alloul:2013fra, Brooijmans:2014eja}.

The vacuum configuration of \eq{eq:desired_vacuum} does not necessarily yield
a minimum of the scalar potential~\cite{Kuchimanchi:1993jg}. In fact, the true
global minimum is in general charge-breaking and the $\Delta_{1R}^{--}$ and
$\Delta_{2R}^{++}$ fields develop a vev
\be\bsp
   \langle \Delta_{iR} \rangle _{\rm{CB}} =
     \begin{pmatrix}
    0 & \frac{v_{iR}}{2} \\
    \frac{v_{iR}}{2} & 0
    \end{pmatrix}
  \ \ \text{for}\ \ i = 1,2\ ,
\label{eq:falseVacuum}
\esp\ee
so that the $D$-term contributions to the scalar potential are minimized. We
will further address this issue in \SEC{sec:hpp}.

In order to design viable left-right supersymmetric benchmark scenarios, we have
implemented the model described above in the \sarah package~\cite{
Staub:2008uz,Staub:2009bi,Staub:2010jh,Staub:2012pb,Staub:2013tta}
to automatically generate a numerical code based on the \spheno
programme~\cite{Porod:2003um,Porod:2011nf}\footnote{The code can be obtained from
the authors upon request.}. Physical spectra are then calculated in several steps.
One begins with the derivation of
the soft-supersymmetry breaking masses from the tadpole equations at the tree
level, which allows for the calculation of a tree-level mass spectrum.
Next, \oneloop corrections are evaluated in the
$\overline{\rm{DR}}$ scheme and a \oneloop-accurate mass spectrum is computed.
As explained in the next section, we have modified the automated
\sarah-\spheno procedure due to special features associated with
the mass of the doubly-charged Higgs
boson. Several analytical cross checks with the model implementation in
\feynrules~\cite{Duhr:2011se,Alloul:2013bka} have been performed.
For the evaluation of the LHC bounds performed in \SEC{sec:LHCbounds},
the \ssp package~\cite{Staub:2011dp,Staub:2015kfa} has been used to scan randomly the
parameter space of the model. Moreover, we have estimated
$W_R$ signal cross sections by means of the \madgraph
programme~\cite{Alwall:2014hca} after having linked the UFO
library~\cite{Degrande:2011ua} obtained from the \sarah implementation.


\section{The doubly-charged Higgs mass}
\label{sec:hpp}

In the class of supersymmetric left-right symmetric models under consideration,
we have four doubly-charged Higgs bosons, two of them being related to the $SU(2)_L$
sector and two of them being related to the $SU(2)_R$ one. While at the tree level,
these fields do not mix, a small mixing is induced at the loop level. It however
does not lead to any observable effect so that the following discussion is
solely focusing on the $SU(2)_R$ sector. The mixing effects are nevertheless
included in our numerical results.

\subsection{Tree-level results}
 \begin{figure}
 \centering
 \includegraphics[width=.49\textwidth]{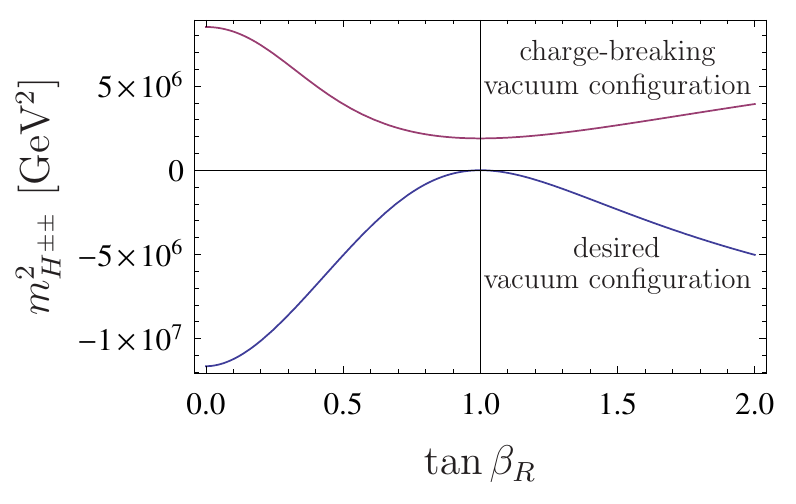}
 \includegraphics[width=.49\textwidth]{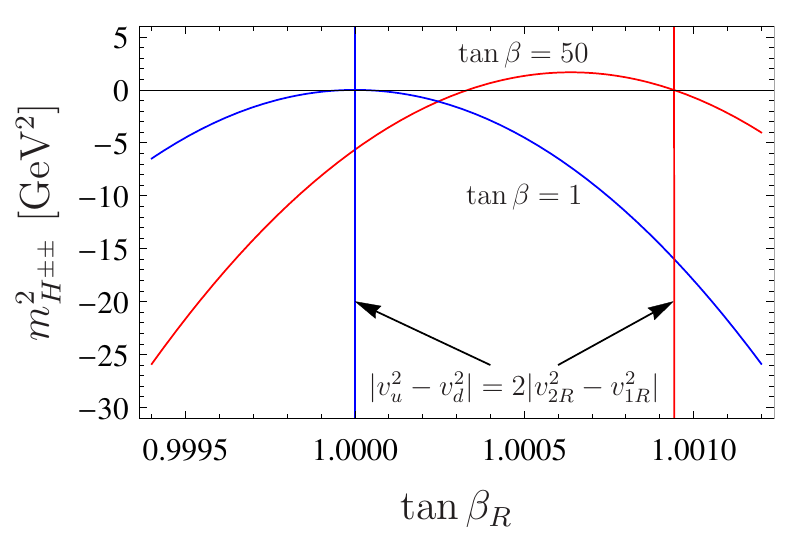}
 \caption{The squared mass eigenvalue of the lightest doubly-charged Higgs state
     $m_{H^{\pm \pm}}^2$ for the desired vacuum configuration of
    \eq{eq:desired_vacuum} (left panel, blue) and for the charge-breaking one of
    \eq{eq:falseVacuum} (left panel, purple) as a function of $\tan \beta_R$. We
    also zoom into the region with $\tan\beta_R\approx 1$ (right panel) and
    present results for the $\tan \beta = 1$ (blue) and $\tan \beta = 50$
    (red) cases. Moreover, the condition $|v_u^2 - v_d^2|> 2 |\vzweir^2 - \veinsr^2|$ that
    is discussed in the text is illustrated by vertical lines. The relevant model
    parameters have been set to
    $v_R = 5.5$~TeV, $\vs = 10$~TeV, $\lambda_R = 0.5$, $\lambda_S = -0.5$,
    $\lambda_{12} = -0.02$ and $T_{\lambda_R} = 0$.}
  \label{fig:treeLevelMass}
 \end{figure}

The squared mass matrix of the doubly-charged $SU(2)_R$ Higgs bosons reads, at
the tree level and in the $(\Delta_{1R}^{--},~{\Delta_{2R}^{++}}^*)$ basis,
\be
  m^2=
  \begin{pmatrix}
    D_{++} - \tan\beta_R\  F_{++} & F_{++} \\
    F_{++} & - D_{++} - \cot\beta_R\  F_{++}
  \end{pmatrix} \ .
\label{eq:tree_level_mass_matrix}
\ee
In our conventions, the $D$-term and $F$-term contributions are respectively
given by
\be\bsp
D_{++} =&\ \frac{\gR^2}{2} \Big[\vd^2 - \vu^2 + 2 (\vzweir^2 - \veinsr^2) \Big]\,, \\
F_{++} =&\ \frac{\lambda_R^2}{2} \veinsr \vzweir + \frac{\lambda_R \lambda_S}{2} \vs^2 
               - \frac{\lambda_R \lambda_{12}}{2} \vd \vu + \frac{T_{\lambda_R}}{\sqrt{2}} \vs\,,
\esp\ee
where $g_R$ stands for the $SU(2)_R$ gauge coupling.
First, the form of this squared mass matrix implies that one of the physical
states is massless in the gaugeless limit, \textit{i.e.}, when all gauge
couplings are neglected. Next, once the contributions proportional to the gauge
couplings are taken into account, the condition
\be
   |v_u^2 - v_d^2| > 2 |\vzweir^2 - \veinsr^2|
\label{eq:vevcondition}\ee
must be satisfied in order to render both doubly-charged Higgs squared mass
eigenvalues positive~\cite{Chacko:1997cm}. Equivalently, this condition can be
casted as $\tan \beta_R -1 \ll 1$ since we have imposed, for phenomenological
reasons, that $v_R \gg v_u, v_d$ (see \SEC{sec:model})\footnote{The special
case where $\tan\beta_R=1$
corresponds to a saddle point and not to a minimum of the potential.}.

The condition of \eq{eq:vevcondition} turns out to be extremely
fine-tuned, as illustrated in \fig{fig:treeLevelMass} where we depict the
dependence of the smallest mass eigenvalue $m^2_{H^{\pm \pm}}$ on
$\tan \beta_R$. In the left panel of the figure, we present numerical predictions
for the desired vacuum configuration of \eq{eq:desired_vacuum} (blue) as well as
for the charge-breaking one of \eq{eq:falseVacuum} (purple). In contrast to the
latter case where the squared mass is generally positive, there exists only a
small region of the parameter space, featuring $\tan \beta_R \approx 1$, where the
vacuum structure does not break electric charge invariance. In the right panel
of \fig{fig:treeLevelMass}, we zoom into this region and study
scenarios for which $\tan\beta_R\approx 1$. In the case where $\tan\beta=1$
(blue), or in other words when $|v_u^2 - v_d^2| = 0$, the condition of
\eq{eq:vevcondition} imposes that the lightest doubly-charged Higgs boson is
either tachyonic or massless. In contrast, when $\tan\beta$ increases, a small
window where the squared mass of the lightest doubly-charged Higgs boson is
positive opens (red, for the example of
$\tan\beta=50$). However, this configuration implies
that one of the pseudoscalar Higgs bosons gets tachyonic~\cite{Huitu:1994zm}.

As a consequence, it is impossible to construct, at the tree level, left-right
supersymmetric setups that are phenomenologically viable. The only
alternative option would be to consider solutions of the potential
minimization equations featuring $R$-parity breaking vevs for the right-handed
sneutrinos~\cite{Kuchimanchi:1993jg,Huitu:1994zm}.

\subsection{Results at the one-loop level}
\label{sec:loop_corrections}
\begin{figure}
\centering
 \includegraphics[width=.49\linewidth]{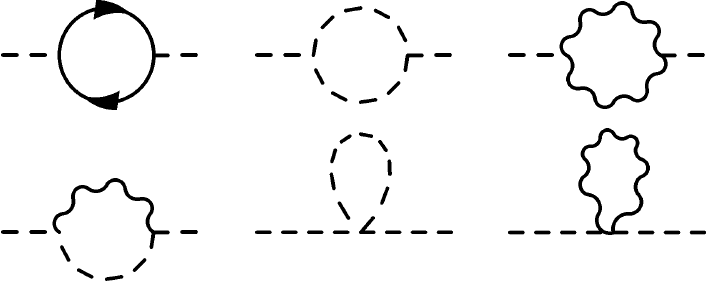}
\caption{Generic \oneloop diagrams contributing to the doubly-charged Higgs self-energy.}
\label{fig:feynman}
\end{figure}

The problems mentioned above are a tree-level artefact and can be solved by
considering one-loop contributions. Attempts including quantum
corrections derived either from the Coleman-Weinberg
potential~\cite{Cvetic:1983su,Babu:2008ep} or within a Feynman diagrammatic
approach~\cite{Babu:2014vba} have shown that a
charge-conserving minimum of the scalar potential could be obtained. A complete
one-loop calculation however exhibits logarithmic terms with a negative argument
resulting from contributions depending on the self-interactions of the
doubly-charged Higgs bosons, and the subsequent imaginary part of the one-loop
corrected potential is a sign of a dangerous unstable vacuum (see
\SEC{sec:vacuum_stability}).
More precisely, these earlier works have argued
that contributions stemming from the Majorana neutrino Yukawa coupling
$y^L_4$ were sufficient to render the doubly-charged Higgs squared mass
positive~\cite{Babu:2014vba} and to make the vacuum configuration of
\eq{eq:desired_vacuum} a deeper minimum than the one of
\eq{eq:falseVacuum}~\cite{Babu:2008ep}.
However, in this last case, the right slepton soft squared masses
$m_{L_R}^2$ must be negative, which further implies large one-loop gaugino
contributions to the slepton mass eigenvalues
to avoid charge and $R$-parity breaking.

\begin{figure}
 \centering
  \includegraphics[width=.49\linewidth]{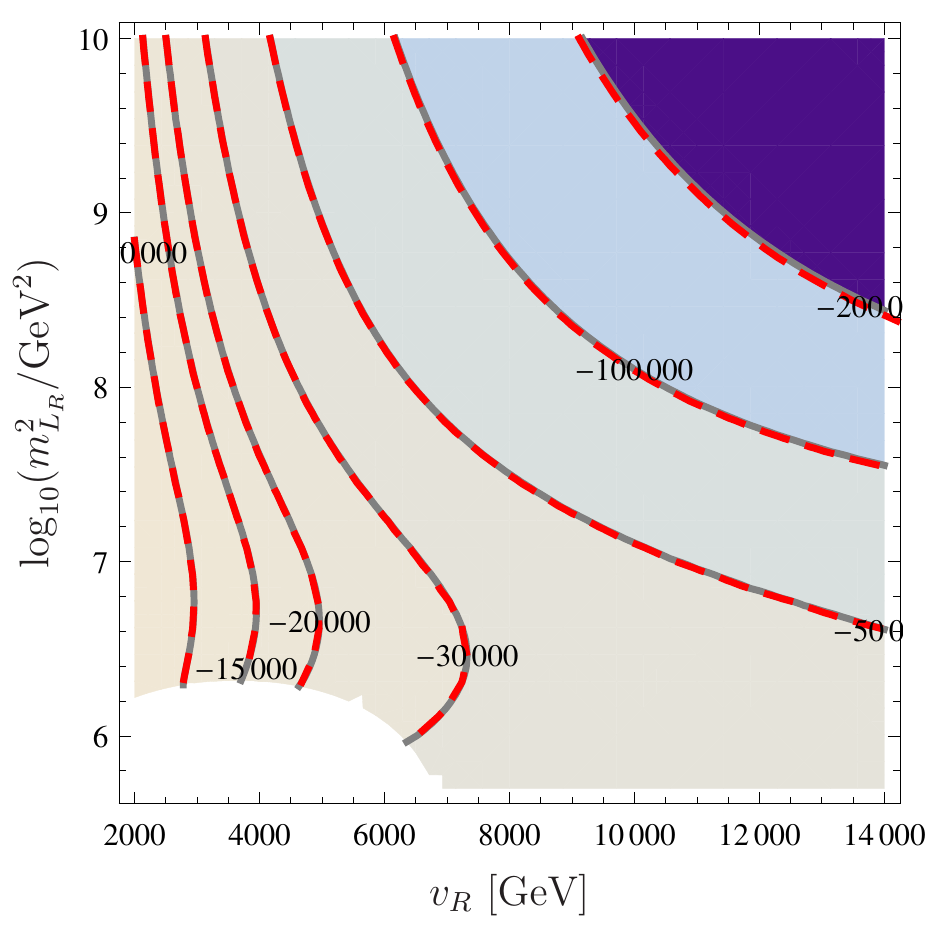}
  \includegraphics[width=.49\linewidth]{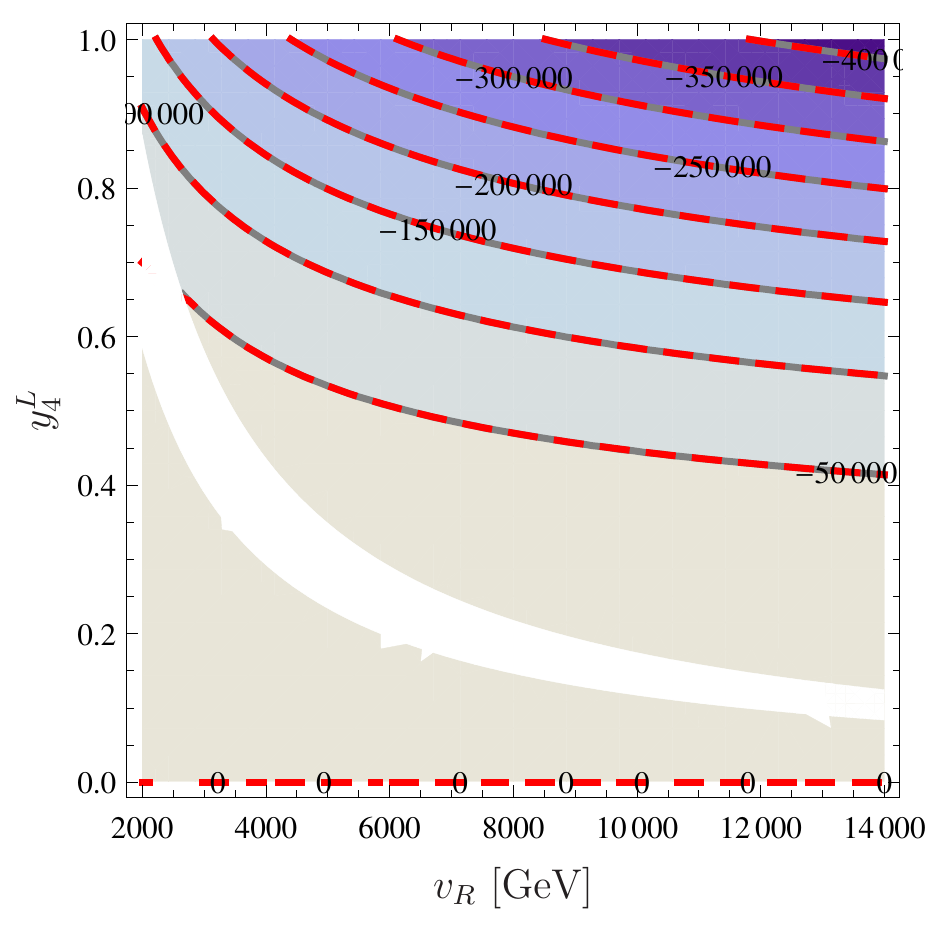}
\caption{Comparison of our calculation of the one-loop corrections from $y^L_4$, evaluated in the
  gaugeless limit and for one generation of right-handed neutrinos (red dashed),
  to the results of \REF{Babu:2014vba} (grey). The contour lines represent
  $m_{H^{\pm \pm}}^2$ isomass lines in GeV$^2$.
  We present results in the $(v_R, \log_{10} (m_{L_R}^2/\rm{GeV}^2))$ plane with
  $y_4^L=0.4$ (left panel) and in the $(v_R, y_4^L)$ plane with
  $m_{L_R}^2 = 2 \cdot 10^6$~GeV$^2$ (right panel). In both cases, the remaining
  relevant parameters are fixed to $\lambda_R = 0.4,~\tan \beta_R=1.02$ and $\vs=10$~TeV and
  the white areas are regions where one of the right sneutrinos is tachyonic.}
 \label{fig:comparisonBabuStyle}
\end{figure}

In this work, we have extended the Feynman-diagram-based calculations of
\REF{Babu:2014vba}, the latter having been achieved in the gaugeless
limit, for one generation of right-handed neutrinos and considered only
diagrams depending on $y^L_4$. We have instead computed all
one-loop contributions to the doubly-charged Higgs boson mass, the associated
Feynman diagrams being presented in \fig{fig:feynman}. In
\fig{fig:comparisonBabuStyle}, we start by numerically comparing our results,
evaluated in the appropriate limit,
to those of \REF{Babu:2014vba} and demonstrate that a good
agreement has been found\footnote{The authors of \REF{Babu:2014vba} have
confirmed a typo in the first line of their eq.~(138) in which the `$+1$' should
be read outside of the logarithm.}. In the parameter space regions
probed on the figure, the soft slepton mass $m_{L_R}^2$ and the effective
supersymmetric Higgs mass parameter
$\mu_R^{\rm eff}=\lambda_R \vs /\sqrt{2}$ are not too large, which
leads to a doubly-charged Higgs boson that is still tachyonic. The loop corrections are even
in this case negative and thus counterproductive to restore a vacuum state that
conserves the electric charge. Only for large values of $\lambda_R \vs$ (not
shown on the figure), a positive contribution could emerge.
This requirement however also yields a significant $CP$-splitting of the right
sneutrinos, so that large values for $m_{L_R}^2$ are as well required to prevent
them from being tachyonic and breaking $R$-parity. Assuming that all sfermions
have similar masses, realistic setups of the class of models under
study are thus unlikely to be observable from standard sfermion
searches at the LHC.

As shown in the rest of this section, the situation improves when the complete
one-loop result is considered. Technically, our predictions rely on the
{\tt SARAH-SPheno} procedure for generating the particle spectrum. In the usual
running mode of both programmes, the tree-level $\overline{\mbox{DR}}$-masses are
used for the particles running into the loop diagrams. In the left-right
supersymmetric models under consideration, this leads to problematic
$\log(m^2_{H^{\pm\pm}}/Q^2)$ terms as the lightest doubly-charged Higgs boson is
tachyonic at the tree level. To avoid this issue, we have modified the
{\tt SARAH-SPheno} procedure in a way that is inspired by the on-shell scheme,
and inserted \oneloop corrected masses into the loops. Iteratively, this
consists of:
\begin{itemize}
  \item first iteration: $m^2_{[1]} = m^2(m^2_{[0]})$ using $ m^2_{[0]}  = |[m^2_{H^{\pm \pm}}]_{\rm tree-level}|$;
  \item  $m^2_{[k]} = m^2(m^2_{[k-1]})$ until $\frac{|m_{[k]} - m_{[k-1]}|}{m_{[k]}} < \epsilon \ll 1$.
\end{itemize}
We have fixed $\epsilon = 10^{-4}$ so that four iterations are generally
necessary for any benchmark scenario.
Moreover, the initial value $m^2_{[0]}$ is in principle arbitrary so that we
could have chosen $m^2_{[0]} = 0$. We have verified the invariance of the
predictions with respect to this choice. Although our prescription breaks gauge
invariance at the two-loop level, the associated effects are expected to be
significantly smaller than the genuine two-loop contributions and thus under
good control. Here and in the following sections $\mhpp$ denotes the
mass of the lightest doubly-charged Higgs boson.

\begin{figure}
\centering
 \includegraphics[width=.49\linewidth]{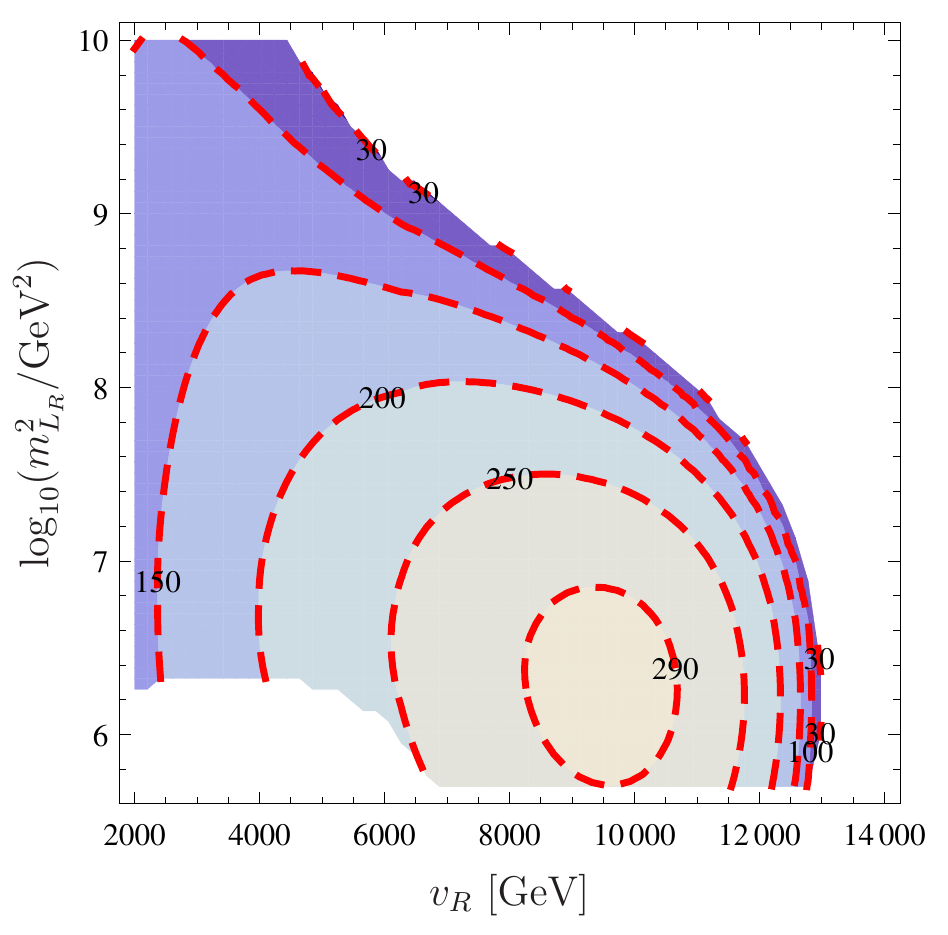}
 \includegraphics[width=.49\linewidth]{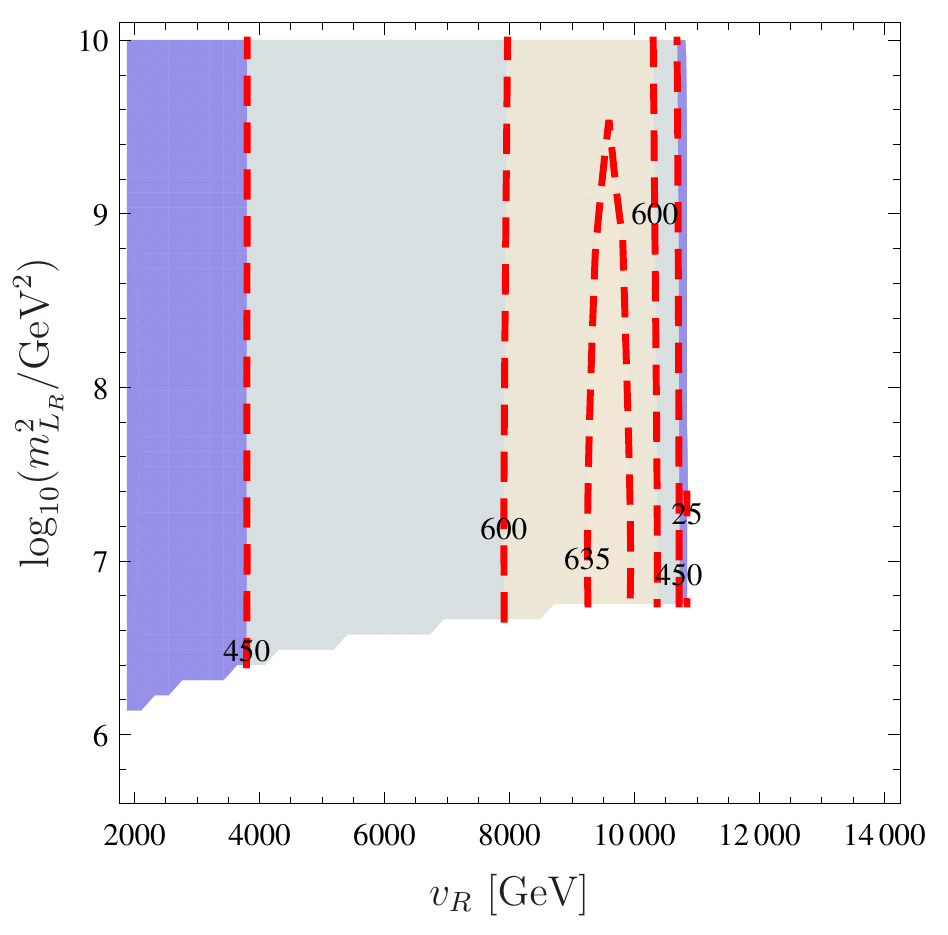}
\caption{Contours of $\mhpp$ [in GeV] including the
 complete one-loop calculation in the
 $(v_R,\log_{10} (m_{L_R}^2/\rm{GeV}^2))$ plane for a scenario featuring the
 same setup as  the left panel of \fig{fig:comparisonBabuStyle} (left), as
 well as for a scenario featuring a smaller $y_4^L=0.1$ and a larger
 $\lambda_R=0.9$ value (right).}
\label{fig:first_full_plot}
\end{figure}

In \fig{fig:first_full_plot}, we show predictions for the \hppmm mass obtained
when using our complete one-loop calculation. The results are presented in
$(v_R,\log_{10} (m_{L_R}^2/\rm{GeV}^2))$ planes, first (left panel of the
figure) when all the other model parameters are fixed as in the left
panel of \fig{fig:comparisonBabuStyle}. Although both the
tree-level and the approximate one-loop predictions yield tachyonic
doubly-charged Higgs bosons, the fully-\oneloop-corrected mass is positive and
of the order of a few hundreds of GeV. Moreover, the dependence on $m_{L_R}^2$
is explained by the $y_4^L$-dependent contributions. In the right panel of
\fig{fig:first_full_plot}, we increase the value of $\lambda_R$ and decrease
the one of $y_4^L$ so that the dependence on $m_{L_R}^2$ almost vanishes
and the one-loop-corrected \hppmm mass reaches values beyond 500~GeV.
The most important contributions to the full one-loop result consist of positive
corrections arising from $W_R$/singly-charged Higgs boson and
doubly-chargino/neutralino loop-diagrams, as well as from
negatively contributing $y_4^L$-induced loop-diagrams,
singly-charginos and doubly-charged Higgs bosons.

\begin{figure}
\centering
 \includegraphics[width=.49\linewidth]{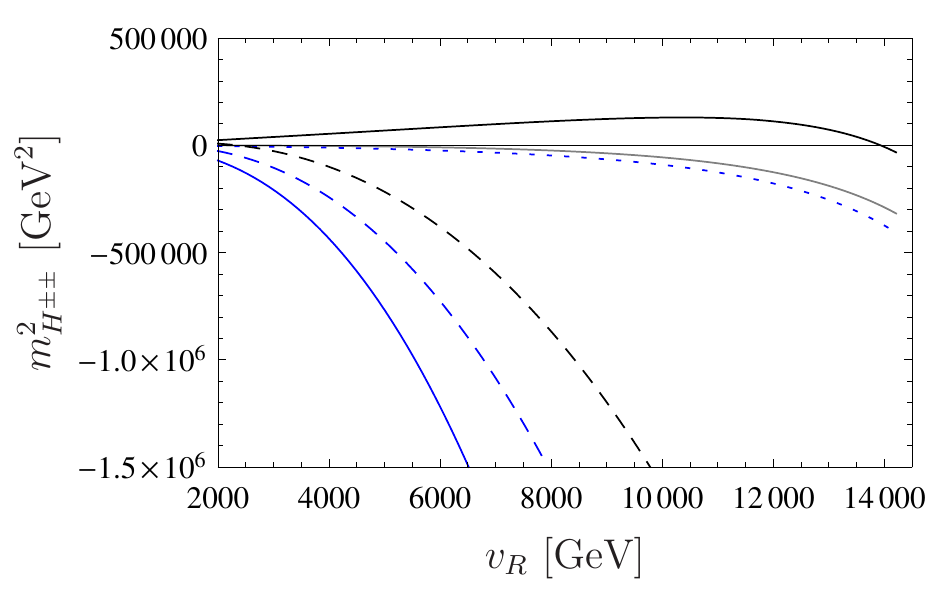}
\caption{Dependence of $\mhpp^2$  on the $SU(2)_R$-breaking scale $v_R$. The lines correspond to
  to the tree level prediction (grey solid) and  add then successively different
  contributions: tree-level +  $y_4^L$-dependent (s)lepton and (s)neutrino contributions (blue dotted),
  +  neutral/doubly charged Higgses and neutral gauge bosons (blue
  dashed), +  chargino contributions (blue solid), + neutralinos and doubly-charged Higgsinos 
  (black  dashed). The full result (black solid) finally  contains also  $W_R/H^\pm$
  contributions. The employed benchmark scenario is defined by $\lambda_R=0.4$, $\tan \beta_R = 1.02$, 
  $m_{L_R}^2 = 2\cdot 10^6$~GeV, $y_4^L = 0.25$ and features one generation of
  right-handed neutrinos.}
  \label{fig:evolution_vR}
\end{figure}

The parametric dependence of the loop contributions is highly non-trivial because of the 
large number of different contributing sectors. However, some generic features can be extracted
using the information given in \fig{fig:evolution_vR}, where we add successively different 
loop contributions for an exemplary benchmark scenario. We start by adding to the tree-level
results the $y_4^L$-dependent contributions which have already been discussed: they become negative
for large values of  $y_4^L$ and $m^2_{L_R}$.

In order to describe the additional contributions
that will be further added, we recall that the structure of the relevant chargino and neutralino couplings
reads
\begin{align}
&\Gamma^L_{\tilde{\chi}_R^{++}\tilde{\chi}^0_{{j}}H^{--}_{{k}}}  &=  &-i \Big( \lambda_R U^{0,*}_{j, \tilde S} 
Z_{{k, \Delta_{1R}^{--}}}^{--} 
 +\sqrt{2} \Big(g_{BL} U^{0,*}_{j, \tilde B} 
 Z_{{k, \Delta_{2R}^{++,*}}}^{--}  + g_R U^{0,*}_{j, \tilde W_{R,3}} 
 Z_{{k, \Delta_{2R}^{++,*}}}^{--} \Big)\Big)\,,\nonumber\\ 
&\Gamma^R_{\tilde{\chi}_R^{++}\tilde{\chi}^0_{{j}}H^{--}_{{k}}}  &=  &-i \Big(  \lambda_R U_{{j, \tilde S}}^{0} Z_{{k, \Delta_{2R}^{++,*}}}^{--} 
  -\sqrt{2} \Big( g_{BL} U_{{j, \tilde B}}^{0} Z_{{k, \Delta_{1R}^{--}}}^{--}  + g_R U_{{j, \tilde W_{R,3}}}^{0} Z_{{k, \Delta_{1R}^{--}}}^{--} 
  \Big) \Big)\,,\label{eq:hppmm_neutralino_vertex}\\
&\Gamma^L_{\tilde{\chi}^+_{{i}}\tilde{\chi}^+_{{j}}H^{--}_{{k}}}  &=  &~i \sqrt{2} g_R Z_{{k, \Delta_{2R}^{++,*}}}^{--} \Big( U^{+,*}_{i,\tilde W_R^+} U^{+,*}_{j,\tilde \Delta^+_R}   +  U^{+,*}_{i,\tilde \Delta^+_R} U^{+,*}_{j,\tilde W_R^+}  \Big)\,,\nonumber \\ 
&\Gamma^R_{\tilde{\chi}^+_{{i}}\tilde{\chi}^+_{{j}}H^{--}_{{k}}}  &=  &-i \sqrt{2} g_R Z_{{k,\Delta_{1R}^{--}}}^{--} \Big( U_{{i,\tilde W_R^+}}^{-} U_{{j,\tilde \Delta^+_R}}^{-}  + U_{{i,\tilde \Delta^+_R}}^{-} U_{{j,\tilde W_R^+}}^{-} \Big)\,, \nonumber
\end{align}
where we have parameterised the vertices as $\Gamma = \Gamma^L P_L + \Gamma^R P_R$.
In our notations, $U^0,~U^\pm$ and $Z^{--}$ are the mixing matrices of the neutralinos, charginos and doubly-charged Higgs
bosons respectively, whose elements are written as $U_{i,X}$ and $Z_{i,X}$. These matrix elements
hence denote the $X$ field component of
the $i^{\rm th}$ mass eigenstate. We also introduce a set of vector and Higgs boson interactions whose strengths are
given by
\begin{align}
&\Gamma_{H^{--}_{{j}}W_R^{+} W_R^{+}}  &= & 
-i \sqrt{2} g_{R}^{2} \Big(  v_{1R}  Z_{{j, \Delta_{1R}^{--}}}^{--}  + v_{2R}  Z_{{j, \Delta_{2R}^{++}}}^{--} \Big)\,, \nonumber \\
&\Gamma_{H^{--}_{{j}}H^+_{{k}}W_R^{+}}  &= & \, 
-i \Big(g_R   Z_{{k,\Delta_{1R}^{-,*}}}^{-} Z_{{j, \Delta_{1R}^{--}}}^{--}  + g_R   Z_{{k,\Delta_{2R}^+}}^{-} Z_{{j, \Delta_{2R}^{++,*}}}^{--} \Big)\,,\\ 
&\Gamma_{H^{--}_{{i}}H^+_{{j}}H^+_{{k}}}  &= &~\frac{i}{\sqrt{2}} g_{R}^{2} \Big(2   \left(v_{2R} Z_{{j,\Delta_{2R}^{+}}}^{-} Z_{{k,\Delta_{2R}^{+}}}^{-} Z_{{i, \Delta_{2R}^{++,*}}}^{--} + v_{1R} Z_{{j,\Delta_{1R}^{-,*}}}^{-}   Z_{{k,\Delta_{1R}^{-,*}}}^{-} Z_{{i, \Delta_{1R}^{--}}}^{--}\right) \nonumber \\
& & &-  \left(Z_{{j,\Delta_{2R}^{+}}}^{-} Z_{{k,\Delta_{1R}^{-,*}}}^{-} + Z_{{j,\Delta_{1R}^{-,*}}}^{-} Z_{{k,\Delta_{2R}^{+}}}^{-} \right) \left(v_{2R}  Z_{{i, \Delta_{1R}^{--}}}^{--}  +  v_{1R}  Z_{{i , \Delta_{2R}^{++,*}}}^{--} \right)\Big) \,, \nonumber
\end{align}
the momentum and metric dependence being omitted from the Feynman rules.

Beside the $y_4^L$ contributions, a set of self-energy diagrams also contribute negatively. These involve
either the doubly-charged Higgs boson and a neutral vector or Higgs boson or
two singly-charged charginos. The sign of the latter, where one
would have naively expected a positive contribution, can be understood
from the large mixing between the gaugino and the singly-charged Higgs bosons
of the $SU(2)_R$ sector.
The contributions to the diagonal entries of the doubly-charged Higgs mass matrix are proportional to
$(|\Gamma^L_{\tilde{\chi}^+_{{i}}\tilde{\chi}^+_{{j}}H^{--}_{{k}}}|^2+ 
|\Gamma^R_{\tilde{\chi}^+_{{i}}\tilde{\chi}^+_{{j}}H^{--}_{{k}}}|^2) \mhpp^2$ whereas the contributions to its
off-diagonal entries are proportional to $(\Gamma^L_{\tilde{\chi}^+_{{i}}\tilde{\chi}^+_{{j}}H^{--}_{{k}}})
(\Gamma^R_{\tilde{\chi}^+_{{i}}\tilde{\chi}^+_{{j}}H^{--}_{{l}}})^* m_{\tilde \chi_i^+} m_{\tilde \chi_j^+}$.
The masses of the relevant charginos are however proportional to $v_R$ and the
wino soft supersymmetry-breaking mass so that they are
in general much larger than $\mhpp$. The contributions to the off-diagonal entries of
the doubly-charged Higgs mass matrix turn therefore out to be much larger than those to the diagonal ones. After
diagonalisation, this yields a negative contribution to $\mhpp$.

The main positive contribution to the doubly-charged Higgs mass are loop diagrams
containing a neutralino and a doubly-charged higgsino. The difference
with the singly-chargino case depicted above stems from the singlino component of the
neutralinos. In the respective entries of the mass matrix, the product
$(\Gamma^{L/R}_{\tilde{\chi}_R^{++}\tilde{\chi}^0_{{j}}H^{--}_{{k}}}) 
(\Gamma^{L/R}_{\tilde{\chi}_R^{++}\tilde{\chi}^0_{{j}}H^{--}_{{l}}})^*$ involves $\lambda_R$-dependent
terms whose sign is different from the one of the gauge contributions. These terms
will hence dominate for large values of $\lambda_R$.
Additionally, the $W_R/H^+$ loop diagrams also contribute positively to the diagonal
entries of the $H^{\pm\pm}$ mass matrix.

Eventually, the negative growth of the tree-level contribution 
to the \hppmm mass (grey line in \fig{fig:evolution_vR}) with increasing $v_R$ cannot
be compensated anymore so that
the \hppmm gets again tachyonic for large enough $v_R$. This scale turns
to be $v_R \simeq 14~$TeV for the example of \fig{fig:evolution_vR}.

Both LHC collaborations have set bounds on the mass of the doubly-charged Higgs
boson. These limits however strongly depend on the final state in which the
doubly-charged Higgs boson decays into~\cite{Chatrchyan:2012ya,ATLAS:2014kca}.
In this way,
$m_{H^{\pm\pm}}$ is constrained to be larger than 204~GeV, 459~GeV, 396~GeV and
444~GeV for $\tau\tau$, $\mu\mu$, $e\mu$ or $ee$ final states, respectively. As
shown in \fig{fig:first_full_plot}, there is usually a large range of
$v_R$ values, while keeping all other parameters fixed, for which these LHC
constraints are satisfied. However, $v_R$ also sets the mass scale of the
$W_R$ boson so that one can combine these constraints with LHC
limits on the $W_R$ mass (see \SEC{sec:LHCbounds}).


\section{LHC bounds}
\label{sec:LHCbounds}

The model parameters entering in the above computations can also be
experimentally constrained by searching for the rest of the particle spectrum.
We focus here on the electroweak part, \textit{i.e.}, extra gauge bosons and
right-handed neutrinos, updating the results first given in the report of the
2013 Les Houches workshop~\cite{Brooijmans:2014eja}.

In the left-right supersymmetric setup under study in this work, the extra $Z_R$
boson is always heavier than its $W_R$ charged counterpart. We hence focus on
the latter since it is more constraining. In general, a $W_R$ boson can decay into pairs of fermions as the Standard Model $W$ boson, into pairs of Standard Model bosons, into pairs of sfermions, into a chargino and a neutralino final state, and into a charged lepton and a right-handed neutrino. Therefore, one can experimentally look for it in several search channels, each setting its own bounds. The seemingly most stringent one is typically set by investigating the signature of a charged lepton and missing energy, assuming that the new $W_R$ boson decays into a lepton and a low-mass neutrino which escapes detection. However, in a left-right scenario this bound does not apply. When neutrino data is explained by a seesaw mechanism, the decay of the $W_R$ into a lepton and a low-mass neutrino is generally suppressed by the small neutrino mixing, while the $W_R \to \ell \nu_R$ mode will typically give rise to more complicated final states. Before focusing on the latter channel, more involved due to the unknown right-handed neutrino mass,
we review the simpler searches with hadronic two-body final states, \textit{i.e.}, $W_R \to jj$ and $W_R \to tb$. In the first case, we can reinterpret the CMS inclusive dijet search of \REF{Khachatryan:2015sja}. In the second case, a direct comparison to the CMS search for $W_R$ bosons in the $tb$ channel of \REF{Chatrchyan:2014koa} is in order.
We do not consider the ATLAS counterparts because they are less sensitive.
The above searches do apply to left-right models since the
coupling strength of the $W_R$ boson to a pair of quarks is equal to that of the
Standard Model $W$ boson, given that $g_R = g_L$ holds. Their reinterpretation
is done by comparing the $pp\to W_R \to jj(tb)$ cross sections evaluated with
the \madgraph programme~\cite{Alwall:2014hca} to the excluded cross sections as a
function of the $W_R$ mass. More precisely, our predictions have been
evaluated by convoluting leading-order squared matrix elements with the
{\tt CTEQ6L1} set of parton densities~\cite{Pumplin:2002vw} and include a
$K$-factor of $1.23$ and $1.2$ for $pp\to jj$ and $pp\to tb$, respectively.

\begin{figure}
\centering
 \includegraphics[width=.49\linewidth]{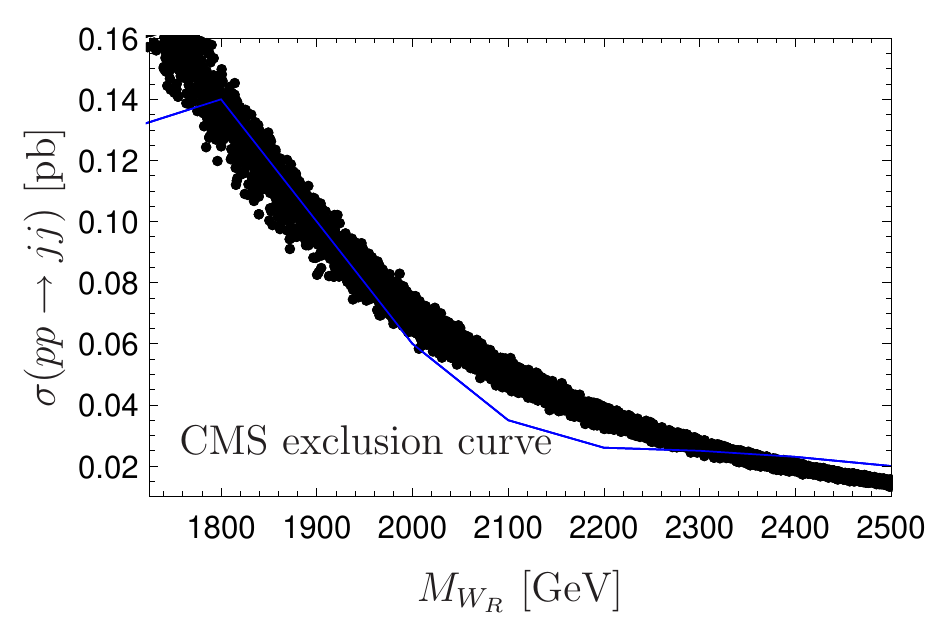}
 \includegraphics[width=.49\linewidth]{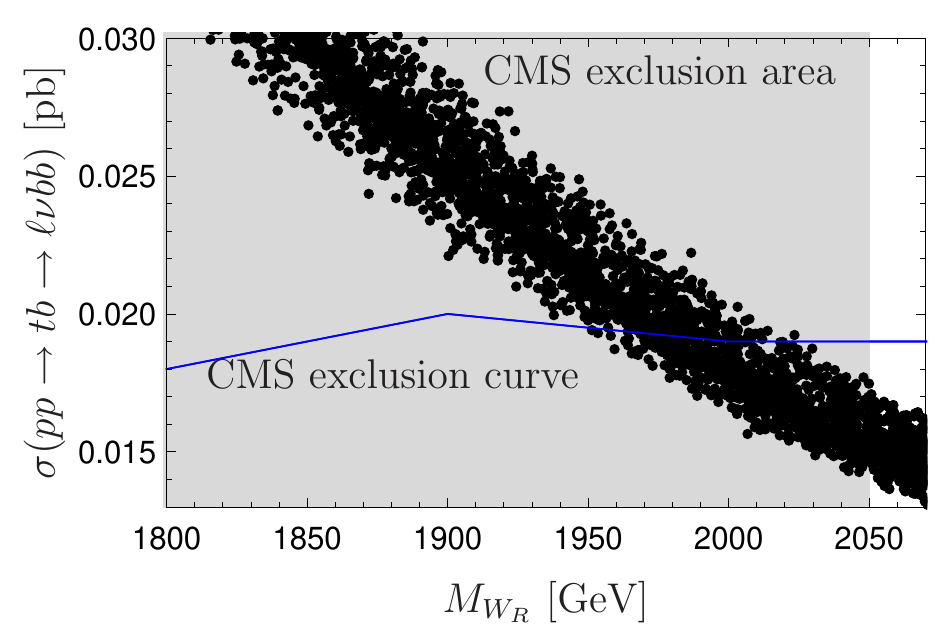}
\caption{Bounds on the $W_R$ boson obtained from the CMS dijet search (left)
  and analysis of $tb$ leptonic events ($\ell = e,\, \mu,\, \tau$) (right). The CMS exclusion curves are taken from
  \REF{Khachatryan:2015sja} and \REF{Chatrchyan:2014koa}, respectively. In the
  latter search, CMS puts a direct bound of $M_{W_R} > 2.05$ TeV, while no such
   bound exists in the former CMS analysis.}
\label{fig:WRbounds_jj-tb}
\end{figure}

In \fig{fig:WRbounds_jj-tb} the cross sections excluded by CMS are compared to those computed in
our model.
In the right panel of the figure, we observe that considering the full model spectrum
decreases the constraint by typically $50-100$ GeV, due to the appearance of
extra decay modes for the $W_R$ that are not considered in the simplified models adopted in the experimental analyses. These extra channels get the $W_R$ decay width proportionally larger, slightly suppressing the branching ratios into the final states that are searched for. The spread in the model prediction reflects the scan over the parameter space. The dijet final state, shown in the left
panel of the figure, is the one setting the tightest bound, 
$M_{W_R} \gsim 2.3$~TeV. The corresponding bound on the right-handed vev is
given by $v_R \gsim 4.9$~TeV. A value of the $W_R$ boson mass between $1.8$ and
$2.0$ TeV, which could be allowed by the dijet search depending on the specific
choice of the input parameters, is fully excluded when the $tb$ search is
also considered.

\begin{figure}
\centering
 \includegraphics[width=.49\linewidth]{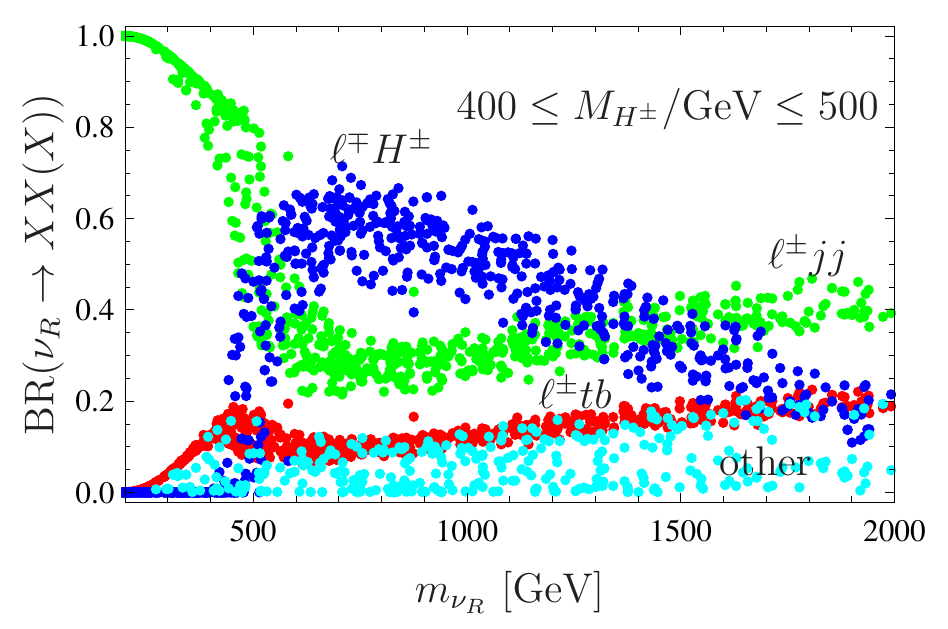}
\caption{Right-handed neutrino branching ratios for a charged-Higgs mass of
  $400~\text{GeV} < M_{H^\pm} < 500$~GeV.
  `Other' refers to all other three-body decays via an off-shell $W_R$ boson.}
\label{fig:hnu-BRs}
\end{figure}

We move now to the constraints in the $W_R \to \ell \nu_R$ process. 
It is usually assumed in left-right symmetric models that the right-handed neutrinos can only decay via the $W_R$ boson, which can be either on- or off-shell, with the subsequent decay of the latter into two jets in
two thirds of the cases. Therefore, footprints of the left-right models are searched for in the $pp\to\ell\ell jj$ ($\ell = e,\, \mu$) final state. This is used by experimental collaborations to set strict bounds on a combination of masses of $\nu_R$ and $W_R$,
as, \textit{e.g.}, in \REF{Khachatryan:2014dka}. As shown in
\REF{BarShalom:2008gt,Brooijmans:2014eja}, this simplified assumption can sometimes be too
restrictive. This is particularly true in the left-right supersymmetric models
under consideration, where new two-body decay channels for the $\nu_R$ can be present, such as $\nu_R \to \ell^\mp H^\pm$ and decays into a neutral (charged) slepton and a chargino (neutralino). Furthermore, several other decay modes for the $W_R$ boson exist, thereby providing new three-body decays for the right-handed neutrinos. This is pictorially shown in \fig{fig:hnu-BRs}.

All these new decay modes for the right-handed neutrinos have the net effect of
reducing its branching ratio into the search channel of interest,
\textit{i.e.}, $\nu_R\to\ell jj$. Two main effects are visible. The right-handed neutrino prefers to decay into the lightest charged Higgs, if this channel is kinematically viable, with the subsequent $H^+ \to t\bar{b}$ decay. However, this decay is possible only due to the (small) right-handed Higgs triplet component of the lightest charged Higgs. Therefore it is dominant only when the $W_R$ needs to be far off-shell for the three-body decays to occur, otherwise the latter will dominate, as happens for large masses
of the right-handed neutrino. We observe the second effect on the figure. The presence of extra decay modes for the $W_R$ (mainly charginos and neutralinos) further suppresses the searched channel, which in the experimental approximation has a constant branching ratio of $67\%$.
However, the branching ratio BR$(\nu_R\to\ell^\mp H^\pm)$ depends on the
right-handed neutrino mass as well as on the singly-charged Higgs mass. It can
get up to $\mathcal{O}(80\%)$ for very light singly-charged Higgs bosons and
right-handed neutrinos and conversely heavy $W_R$ bosons.

\begin{figure}
\centering
 \includegraphics[width=.49\linewidth]{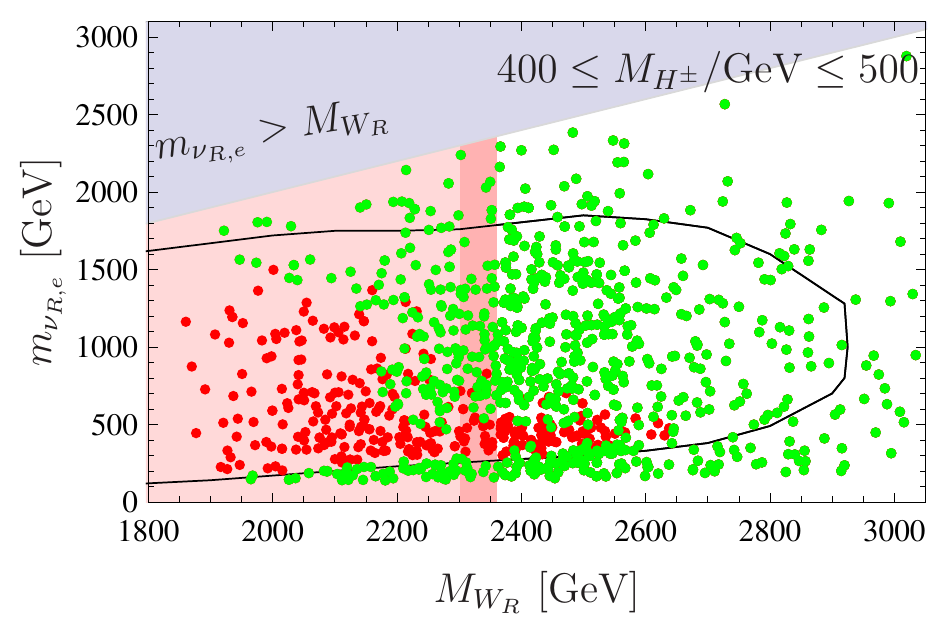}
 \includegraphics[width=.49\linewidth]{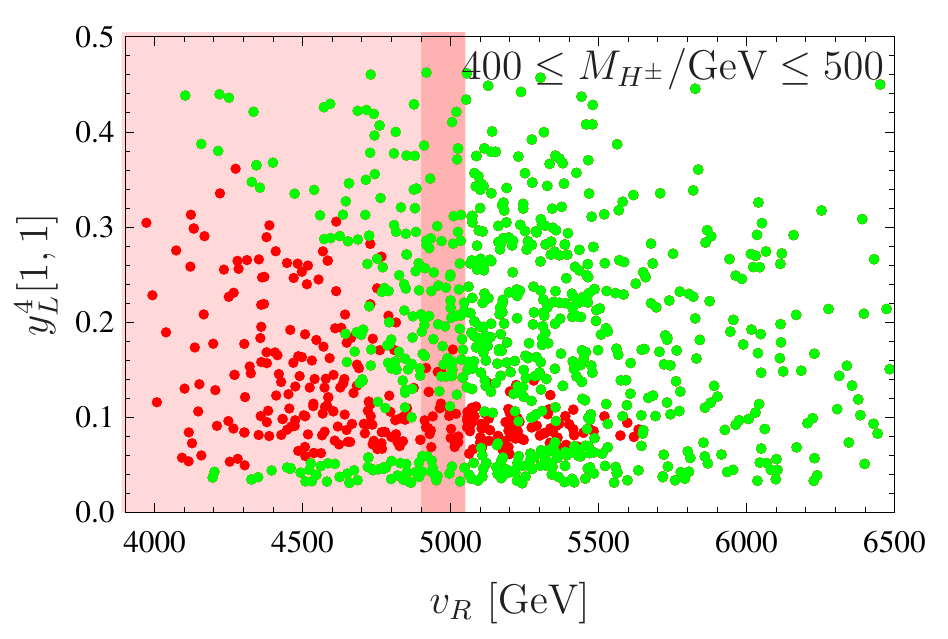}
\caption{Allowed (green) and excluded (red) points from the CMS search for $W_R$
  and $\nu_R$ in the $eejj$ channel for a lightest right-handed neutrino of
  electron flavour (of mass $m_{\nu_{R,e}}$), for $400 < M_{H^\pm}/\mbox{GeV} < 500$, in the
  $(m_{\nu_{Re}}, M_{W_R})$ plane (left) and $(y^4_L[1,1], v_R)$ plane (right).
  The CMS exclusion curve (in black in the left panel of the figure) is taken
  from \REF{Khachatryan:2014dka}. The red shaded area represents the dijet bound.}
\label{fig:mNe-bounds}
\end{figure}

\begin{figure}
\centering
 \includegraphics[width=.49\linewidth]{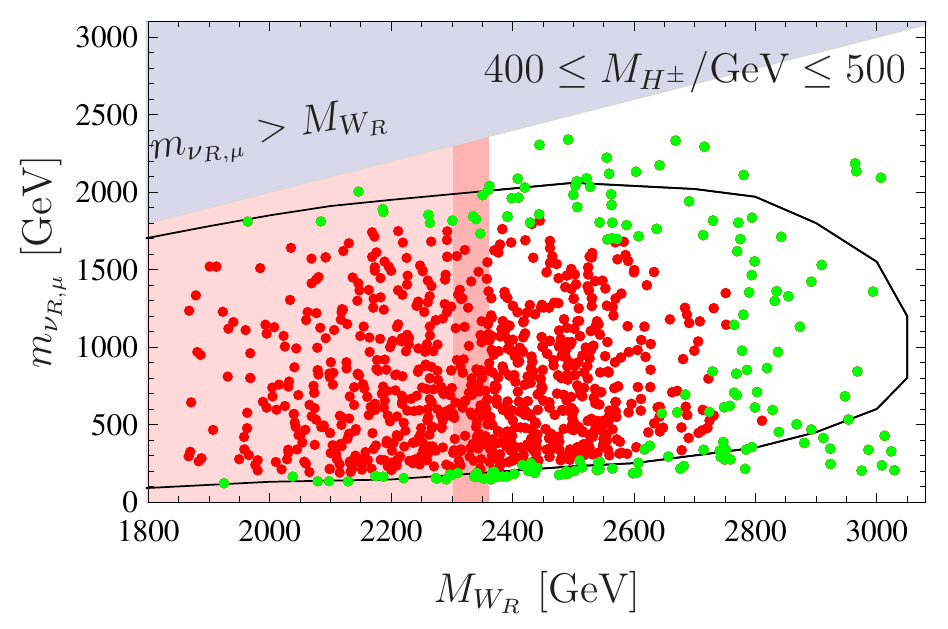}
 \includegraphics[width=.49\linewidth]{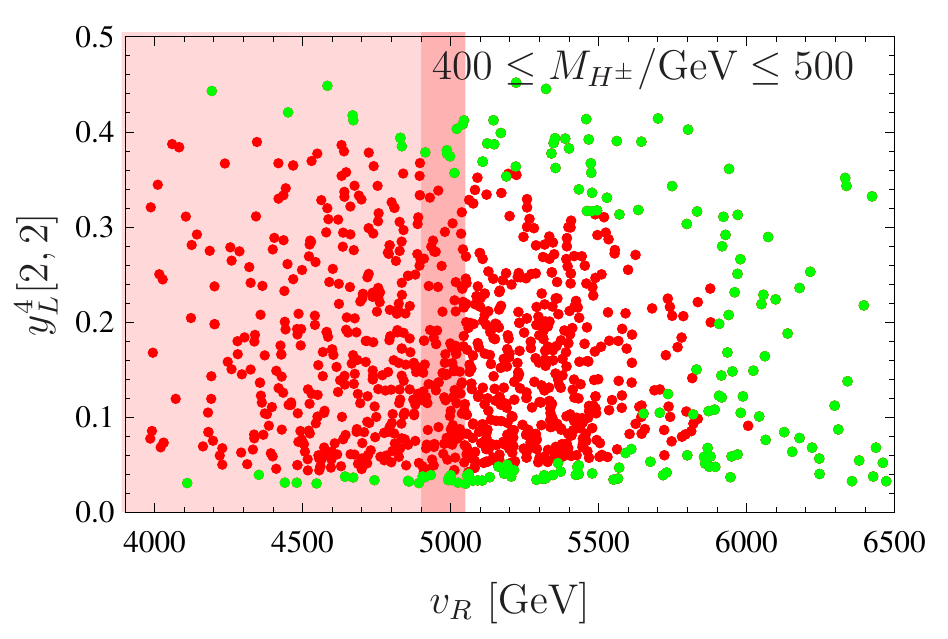}
\caption{Same as in \fig{fig:mNe-bounds}, for a lightest right-handed neutrino of a muon flavour
(of mass $m_{\nu_{R,\mu}}$).}
\label{fig:mNnu-bounds}
\end{figure}

As done previously for the hadronic two-body decays of the $W_R$ boson, we now reinterpret the CMS bounds on the $W_R \to \ell \nu_R \to \ell \ell jj$ channel of \REF{Khachatryan:2014dka}. \Figs{fig:mNe-bounds}--\ref{fig:mNnu-bounds} display the allowed points (green dots) as well as the excluded points (red dots) when the considered right-handed neutrino is of electron and of muon flavour, respectively,
in the $(m_{\nu_R},M_{W_R})$ plane on the left panel of the figure, and in the $(y^4_L,v_R)$ plane
on its right panel.

We have found that the bounds set by CMS are generally too strong to be naively
imposed in our left-right supersymmetric setup, and quite large areas within the
excluded regions survive, especially for an electron right-handed neutrino.
Investigating also \fig{fig:hnu-BRs}, we identify three regions where exist parameter configurations which are allowed. First, for very light right-handed neutrinos, its decay products are too few energetic and appear below threshold. This consists of the lower region, below the black curve, in \figs{fig:mNe-bounds}--\ref{fig:mNnu-bounds}. Next, the region at high masses for both the $W_R$ and the $\nu_R$, that contains allowed points because of extra two-body decays of the $W_R$ that suppress the search channels.
Last, the bulk area in \fig{fig:mNe-bounds} is now allowed, contrary to the CMS
results, because of the suppression of the associated cross section due to the
$\nu_R \to \ell^\mp H^\pm$ decay channel which is dominant for moderate $\nu_R$
masses when such channel is kinematically open. Such an area does not appear for
a muonic right-handed neutrino because of the much stronger CMS bounds.
In fact, a fluctuation of about $3\sigma$ has been found in the $eejj$ CMS search
channel,
for $M_{W_R} \simeq 2.0$ TeV, which our model can fit, as noted in \REFS{Deppisch:2014qpa,Heikinheimo:2014tba} for similar cases.


 \section{Analysis of the vacuum stability}
\label{sec:vacuum_stability}

In \SEC{sec:hpp} we have shown that loop corrections have the power to switch the
$SU(2)_L \times U(1)_Y$ breaking saddle point into a minimum and to give a positive squared mass 
for  \hppmm.  Having now a local minimum with a proper breaking of $SU(2)_L \times U(1)_Y$
does not necessarily imply that this is also the global minimum of the theory.
In particular, it is not obvious that it is deeper lying than the 
charge-breaking minimum of \eq{eq:falseVacuum} after taking into account consistently all
loop effects. 
For further references, we denote in the following
the vacuum configuration with the desired symmetry breaking
pattern as `DSB' and the charge-breaking vacuum as `CB'.

The tadpole conditions cannot be solved analytically at the one-loop level so that one has
to rely on numerical methods. However, the main features of the vacuum configuration
can already be understood by considering a simplified
model where we focus on the $SU(2)_R \times U(1)_{B-L}$ gauge sector, including only 
the two $SU(2)_R$ Higgs triplets $\Delta_{1R}$
and $\Delta_{2R}$ as well as one generation of right-handed (s)leptons.
The corresponding superpotential is given by
\begin{align}
W  = y^L_4\, L_R^T\, \Delta_{1R}\, \epsilon\, L_R 
   + \lamR  \,   S\, {\rm Tr} (\Delta_{1R} \,\Delta_{2R})
   + \frac13 \lamS\, S^3\,,
\end{align}
where we have written down the full $SU(2)$ matrix structure
for the first term using \mbox{$L_R = (\nu_R^c,\ell_R^c)^T$} as well as the $\epsilon$ tensor
defined by $\epsilon_{12}=-\epsilon_{21}=1$.
The tree-level scalar potential $V_0$ reads
\be
V_0 = V_F + V_D + V_{\rm soft}
\ee
with
\be\bsp
V_F &= \lambda_R^2\, |S|^2 \, \Big( {\rm{Tr}} (\Delta_{1R}^\dagger \, \Delta_{1R}) + {\rm{Tr}} (\Delta_{2R}^\dagger \, \Delta_{2R}) \Big) + \Big| \lambda_R \,{\rm{Tr}} (\Delta_{1R}\,  \Delta_{2R}) + \lambda_S\, S^2 \Big|^2 \\
&\quad +y^L_4 \,\lambda_R \left( S^*\, \tilde L_R^T\, \Delta_{2R}^\dagger\, \epsilon\, \tilde L_R +h.c. \right)
+ (y^L_4)^2\, (\tilde L_R^\dagger \tilde L_R)^2\\
V_D &= \frac{g_R^2}{2} \sum_{i=1}^3 \Big( {\rm Tr} (\Delta_{1R}^\dagger [\tau_i, \Delta_{1R}])+
 {\rm Tr} (\Delta_{2R}^\dagger [\tau_i, \Delta_{2R}]) -\tilde L_R^\dagger\, \tau_i\, \tilde L_R\Big)^2 \\&\quad + 
 \frac{g_{BL}^2}{2} \Big( {\rm Tr}(\Delta_{2R}^\dagger\, \Delta_{2R}) - {\rm Tr}(\Delta_{1R}^\dagger\, \Delta_{1R}) +\frac{1}{2} \tilde L_R^\dagger  \tilde L_R\Big)\\
V_{\rm soft}  &= m^2_{\Delta_{1R}} {\rm Tr}(\Delta_{1R} \, \Delta_{1R}^\dagger) + m^2_{\Delta_{2R}} {\rm Tr}(\Delta_{2R}\,  \Delta_{2R}^\dagger) + m_S^2 |S|^2 + m_{L_R}^2 \tilde L_R^\dagger \tilde L_R \\
  &\quad +\Big( T_{\lambda_R}\, S\, {\rm Tr}(\Delta_{1R}\, \Delta_{2R})
+ \frac{1}{3} T_{\lambda_S}\, S^3 + T^L_4\, \tilde L_R^T\, \Delta_{1R}\, \epsilon\, \tilde L_R + {\rm h.c.} \Big) \,.
\esp\ee
In these expressions, $\tau_i=\frac{1}{2} \sigma_i$, where $\sigma_i$ are the Pauli matrices.
For simplicity we set the trilinear soft supersymmetry-breaking couplings to zero as they do not change
the qualitative features that we are aiming to describe.

We start by removing the soft supersymmetry-breaking masses
$m^2_{\Delta_{1R}}$, $m^2_{\Delta_{2R}}$ and $m_S^2$ from the equations by requiring that
the tadpole equations for the desired charge-conserving case of \eq{eq:desired_vacuum} are solved,
\begin{align}
\frac{\partial V_0}{\partial X} \Big|_{{\rm DSB:} \langle \Delta_{iR}^0 \rangle = \frac{v_{iR}}{\sqrt{2}}, \langle S \rangle = \frac{v_S}{\sqrt{2}}} 
= 0\qquad\text{for}\qquad X= \{ \Delta_{1R}^0, \Delta_{2R}^0,S \}\,,
\end{align}
so that the expression of $V_{\rm soft}$ is optimized for the DSB case.
However, as argued above, the global minimum of $V_0$ consists of a configuration where
the vevs of the triplet fields are aligned
along the $\tau_1$ direction. Taking the effect of the soft supersymmetry-breaking masses into account, 
the magnitudes of $v_{iR}$ (with $i=1,2$) get slightly modified by factors $\alpha_i$ which are close to one,
so that we can relate the vevs derived in the DSB case to those derived in the CB case,
\be\bsp
\langle \Delta_{1R}^0 \rangle|_{\rm CB} = \langle \Delta_{1R}^{--} \rangle|_{\rm CB} = \alpha_1 \frac{v_{1R}}{2}
 \qquad\text{and}\qquad
\langle \Delta_{2R}^0 \rangle|_{\rm CB} = \langle \Delta_{2R}^{++} \rangle|_{\rm CB} = \alpha_2 \frac{v_{2R}}{2} \ .
\esp\ee
Consequently, we trade the tadpole equations for the doubly-charged fields with
\begin{align}
\frac{\partial V_0|_{\rm CB}}{\partial \alpha_i} = 0\,.
\end{align}

For the evaluation of the mass spectrum at the DSB and the CB minima, we split all complex	
scalar fields into their scalar and pseudoscalar components as for the CB minimum, as electromagnetism
is eventually broken.
Taking into account all possible vevs, we rewrite every scalar field $X$ as
 $X = \frac{1}{\sqrt{2}}( v_X + \phi^S_X + i \phi^P_X)$. After rotating out the unphysical
would-be Goldstone bosons in each configuration, we compute a $15\times 15$ scalar mass matrix
\begin{align}
\left(M_S^2\right)_{ij}^{\rm DSB/CB} = \frac{\partial^2 V_0 }{\partial \phi_i \partial \phi_j}\Big|_{\rm DSB/CB}\,.
\end{align}
The fermionic part of the spectrum can be evaluated 
for the $SU(2)_R$ ($U(1)_{B-L}$) gauginos $\tilde W_R^i~(\tilde B)$, Higgsino triplets $\tilde \Delta_{iR}$ as well as the the singlet 
fermion $\tilde S$ and the lepton doublet $L_R$
from the Lagrangian terms
\be\bsp
{\mathcal L}_{\rm mass}^{\rm fermions} = & - \lambda_S \, S\, \tilde S \, \tilde S - \lambda_R\, \Big( \tilde S\, {\rm Tr} (\tilde \Delta_{1R}\, \Delta_{2R}) +
\tilde S\, {\rm Tr} ( \Delta_{1R}\, \tilde \Delta_{2R}) + S\, {\rm Tr} (\tilde \Delta_{1R}\, \tilde \Delta_{2R})\Big) \\
&\quad  -\frac{y^L_4}{2}  \Big( 2 L_R^T\, \Delta_{1R}\, \epsilon \, L_R + \tilde L_R^T\, \tilde \Delta_{1R}\, \epsilon\, L_R + L_R^T\, \tilde \Delta_{1R}\, \epsilon\, \tilde L_R \Big)
\\
&\quad - \sqrt{2}g_R \sum_{i=1}^3 \tilde W_{R,i} \Big( {\rm Tr}(\Delta_{1R}^\dagger [\tau_i, \tilde \Delta_{1R}]) +
{\rm Tr}(\Delta_{2R}^\dagger [\tau_i, \tilde \Delta_{2R}]) - \tilde L_R^\dagger \, \tau_i\, L_R \Big)  \\ 
&\quad - \sqrt{2} g_{BL} \tilde B \Big( {\rm Tr} (\Delta_{2R}^\dagger\, \tilde \Delta_{2R}) - {\rm Tr} (\Delta_{1R}^\dagger\, \tilde \Delta_{1R}) +\frac{1}{2} \tilde L_R^\dagger \, \tau_i\, L_R \Big)  \\
&\quad  - \frac{1}{2} m_{\tilde W_R} \sum_{i=1}^3 \tilde W_{R,i}\, \tilde W_{R,i} 
   - \frac{1}{2} m_{\tilde B}\, \tilde B\, \tilde B + {\rm h.c.}\,.
\esp\ee
As an example for the differences between both vacuum configurations, the lepton masses are given by
\be\bsp
m_{\nu_R}|_{\rm DSB} =  \sqrt{2}\, v_{1R}\, y^L_4, \qquad\qquad &m_{e_R}|_{\rm DSB} = 0\,, \\
m_{\nu_R}|_{\rm CB} =  \alpha_1\, v_{1R}\, y^L_4,\qquad\qquad &m_{e_R}|_{\rm CB} = \alpha_1\, v_{1R}\, y^L_4\,.
\esp \ee
Finally, the masses of the vector bosons are derived from the non-derivative part
of the gauge-invariant kinetic terms of the Higgs bosons,
\be\bsp
{\mathcal L}_{\rm mass}^{\rm vector} = &
{\rm Tr}\Big((  g_R \sum_{a=1}^3 W_{R}^{\mu,a}\, [\Delta_{1R}^\dagger,\tau^a]- g_{BL}\, B^\mu\, \Delta_{1R}^\dagger ) (   g_R \sum_{b=1}^3 W_{R,\mu}^b\, [\tau^b,\Delta_{1R}] - g_{BL}\, B_\mu\, \Delta_{1R} )  \\
&\ \ + (  g_R \sum_{a=1}^3 W_{R}^{\mu,a}\, [\Delta_{2R}^\dagger,\tau^a]+ g_{BL}\, B^\mu\, \Delta_{2R}^\dagger ) (  g_R \sum_{b=1}^3 W_{R,\mu}^b \, [\tau^b,\Delta_{2R}]+ g_{BL}\, B_\mu\, \Delta_{2R} ) \Big)\,.
\esp\ee
This gives three heavy states of masses of $\mathcal O(v_R)$ and one massless state for each vacuum structure.
In the DSB case, the
hypercharge symmetry group remains unbroken by the triplet vevs and the associated boson is thus
massless. In the CB case, the vevs are aligned
along the $\tau_1$ direction so that this generator remains
unbroken ($[\tau_1,\Delta_{iR}]=0$) and the $W_{R,1}$ boson turns out to be massless.
This is a consequence of the fact that a Higgs field in the adjoint representation of an $SU(N)$
group cannot break the rank of this group.

\begin{figure}
\includegraphics[width=.49\linewidth]{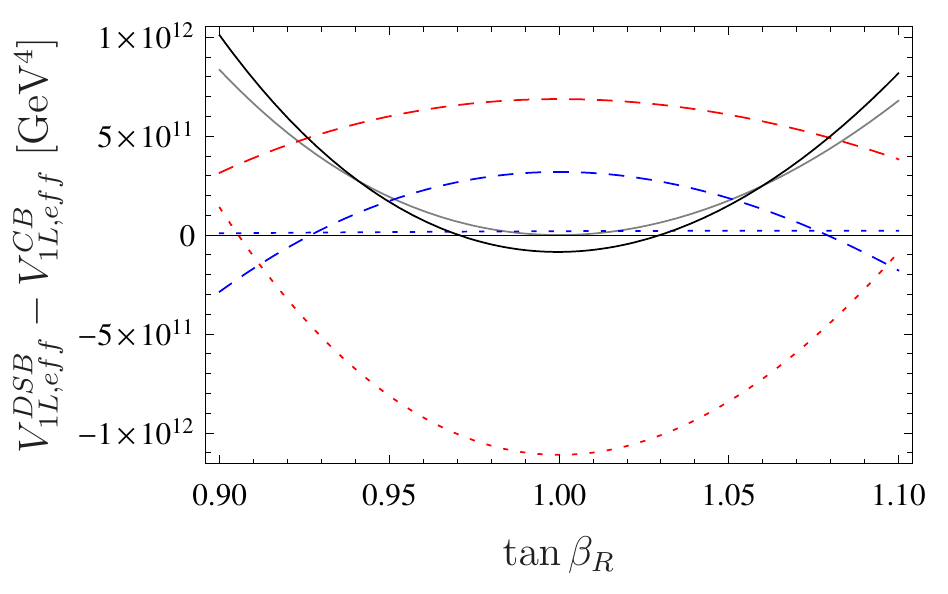}
\includegraphics[width=.49\linewidth]{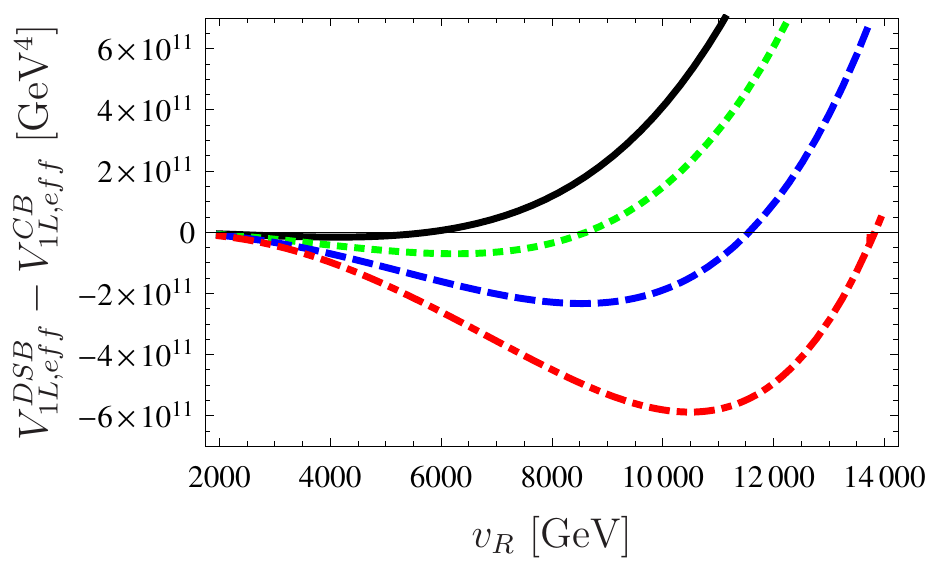}
\caption{Differences of the scalar potential depths between the DSB and the CB cases at the \oneloop level.
For negative (positive) values of this difference, the DSB (CB) minimum is the global one. In the left panel
of the figure, we represent
this difference at the tree (solid grey line) and at the one-loop (solid black line) level as a
function of $\tan \beta_R$ for 
$v_R = 5.5~$TeV, $\lambda_R = 0.4$, $y^L_4=0.25$, $m^2_{L_R} = 2\cdot 10^6~{\rm GeV}^2$ and $v_S = 10~$TeV.
In addition we split the results in terms of the various contributions: slepton and lepton fields (blue dotted line),
Higgs fields (blue dashed line), gauginos/Higgsinos (red dotted line) and
vector bosons (red dashed line).
In the right panel of the figure, we depict the total difference for $\tan \beta_R=1.02$, $y^L_4=0$ and
$\lambda_R = 0.3$ (solid black line), 0.4 (green dotted line), 0.5 (blue dashed line) and 0.6 (red dot-dashed line).
}
\label{fig:potentialdiff_analytic}
\end{figure}

We now move on with the analysis of the \oneloop effective potential
\begin{align}
V_{1L,{\rm eff.}} = V_0 + V_{CW}\,,
\end{align}
where $V_{CW}$ is the Coleman-Weinberg potential. It reads, in the $\overline{\rm DR}$ scheme and using
Landau gauge,
\begin{align}
V_{CW} =  \sum_{n}  \frac{(-1)^{2 s_n} (2 s_n+1)}{64 \pi^2} m_n^4 \Big(\log \left( \frac{m_n^2}{Q^2} \right)  -\frac{3}{2}\Big)\,,
\label{eq:coleman_weinberg}
\end{align}
where $n$ runs over all real scalar fields, Weyl fermions and gauge bosons. We denote by $s_n$ and $m_n$ the respective
spin and mass of the $n^{\rm th}$ field and have also set the renormalization scale $Q$ to
1~TeV.
We show on the left panel of \fig{fig:potentialdiff_analytic} the difference
\be
 \Delta V = V_{1L,{\rm eff.}}^{\rm DSB} - V_{1L,{\rm eff.}}^{\rm CB}
\ee
at the tree-level, at the one-loop level as well as after breaking down the result
for the different contributions to the potential.
For the chosen set of parameters, the global minimum is of the DSB kind up for $\tan \beta_R$ values in the
[0.97, 1.03] range.
As illustrated on this figure, the observed behaviour is a consequence of the interplay between
the fermionic and the bosonic contributions in the Higgs and gauge sectors. In particular,
the DSB vacuum can be the global minimum only due to
the fermionic contributions.
This may seem to contradict the results of the previous section where the charged Higgs and $W_R$
diagrams are  very important for getting a non-tachyonic doubly-charged Higgs. We however recall that
we are focusing here on the differences between the minima and not on the absolute contributions.

We have shown in \SEC{sec:hpp} that a sizeable $\lambda_R$ is needed to get $\mhpp^2$ positive.
On the right panel of \fig{fig:potentialdiff_analytic}, we study the dependence of
$\Delta V$ on $v_R$ for different values of $\lambda_R$.
We observe that increasing $\lambda_R$ also increases the
relative depth of the DSB vacuum with respect to the charge-breaking one.
We have finally checked that the features discussed so far do not depend on the choice
of $Q$ by varying it up to $2 v_R$ and evolving the model parameters according to
the renormalization group equations.


We now turn to the study of the full
model and allow for additional vevs. We have used \sarah to produce a \vevacious
code~\cite{Camargo-Molina:2013qva} for the left-right supersymmetric models under
study. The \vevacious programme starts by evaluating the tree-level scalar potential for
a given spectrum and finds all extrema in terms of all scalar particle vevs, given together with 
the corresponding value of the potential. In a second step, the nearby extrema
are found and evaluated using a full one-loop potential.
In principle one should allow for all scalars to obtain a vev which
however in models like the one considered in this work would take an enormous amount of CPU time 
well beyond a year for a single point of parameter space.
Therefore, we restrict ourselves to vacuum configurations in which
only the $\Phi_{1,2}^0$, $\Delta^0_{1R,2R}$ and $S$ neutral fields and the
$\Delta_{1R}^{--} $ and $\Delta_{2R}^{++}$ doubly-charged fields
could develop non-vanishing vacuum expectation values.
We further allow for one generation of right sneutrinos to receive a vev in order to cover the
possibility of a charge-conserving but $R$-parity violating vacuum.

Under these assumptions, in the parameter regions where an $R$-parity conserving spectrum can be found at the \oneloop level, the global minimum is found to be always either
of the desired kind of \eq{eq:desired_vacuum}, such that
\begin{align}
  \langle \Phi_1^0 \rangle &= &v_d/\sqrt{2}\,, \qquad \quad
  \langle \Phi_2^0 \rangle &= &v_u/\sqrt{2}\,, \qquad
  \langle S \rangle &= &\vs/\sqrt{2} \,, \qquad
  \langle \tilde \nu^c \rangle &= &0 \,,\\
  \langle \Delta^0_{1R} \rangle &= &v_{1R}/\sqrt{2}\,,\qquad
  \langle \Delta_{1R}^{--} \rangle &= &0\,, \qquad \quad
  \langle \Delta^0_{2R} \rangle &= &v_{2R}/\sqrt{2}\,,\quad
  \langle \Delta_{2R}^{++} \rangle &= &0\,,
\end{align}
or the charge-breaking nature of \eq{eq:falseVacuum} with
\begin{align}
  \langle \Phi_1^0 \rangle &= \langle \Phi_2^0 \rangle = 0\,,\qquad \qquad \qquad
  \langle S \rangle = \vs/\sqrt{2} \,,\qquad \qquad \qquad
  \langle \tilde \nu^c \rangle = 0\,,\\
  \langle \Delta^0_{1R} \rangle &= \langle \Delta^{--}_{1R} \rangle \simeq v_{1R}/2\,, \qquad
  \langle \Delta_{2R}^{0} \rangle = \langle \Delta_{2R}^{++} \rangle \simeq v_{2R}/2\,.
\end{align}

\begin{figure}
\centering
 \includegraphics[width=.49\linewidth]{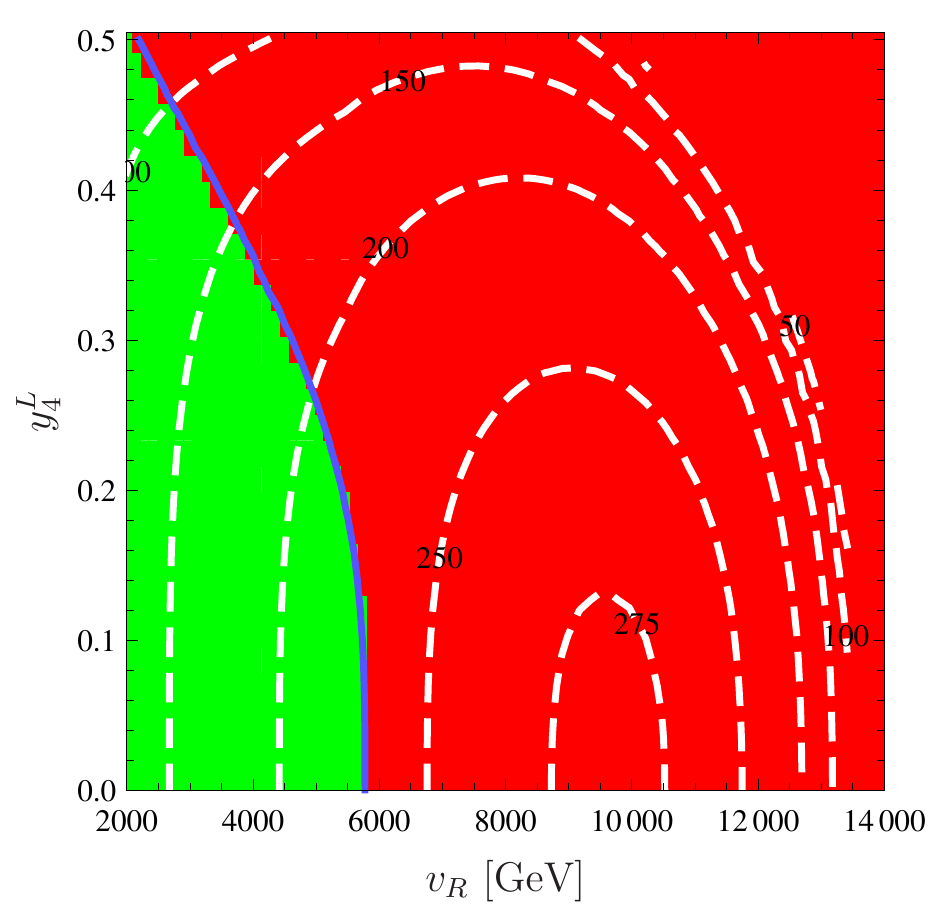}
 \includegraphics[width=.49\linewidth]{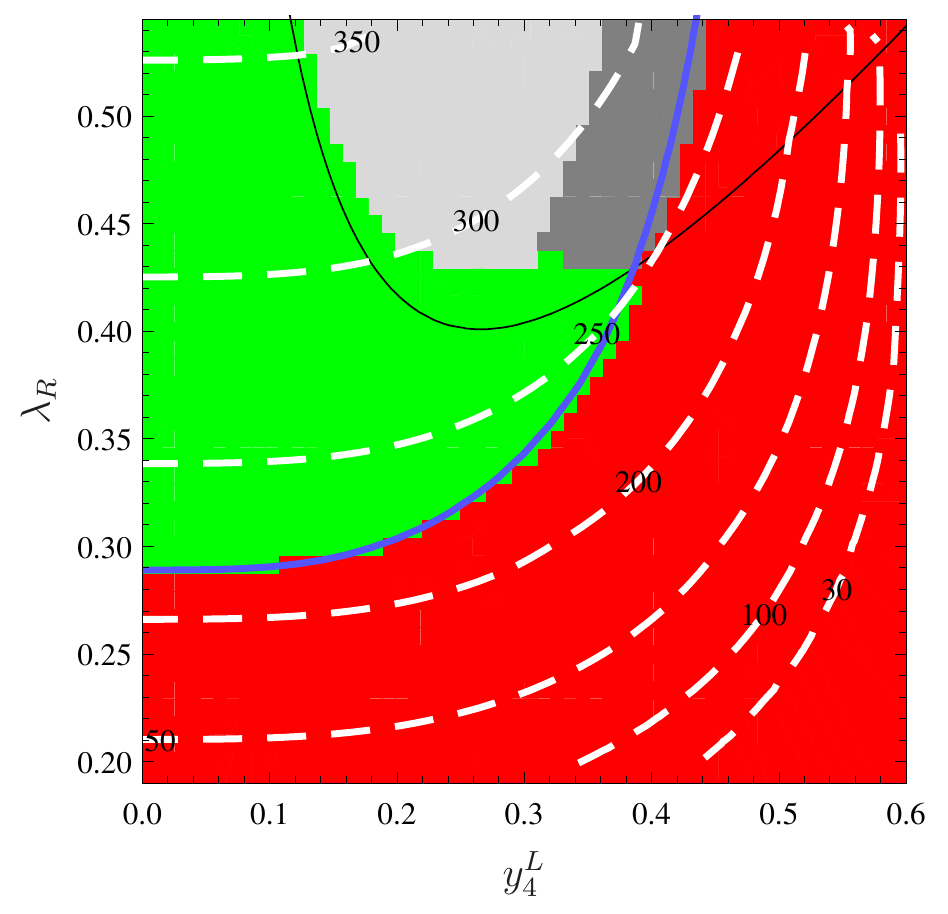}
\caption{Analysis of the vacuum stability. The results are presented in the
  $(v_R,y_4^L)$ plane with  $\lambda_R = 0.3$ and for one generation of
  right-handed neutrinos (left) and in the $(y_4^L,\lambda_R)$ plane with
  $v_R=5.5~$TeV (right). The other model parameters have been fixed to
  $\tan \beta_R=1.02,~m_{L_R}^2 = 2\cdot 10^6~\rm{GeV}^2$ and $\vs = 10~$TeV.
  The white contours depicted on the figures indicate isomass lines for
  the doubly-charged Higgs boson in GeV as obtained with our full one-loop-corrected
  calculation. The green regions correspond to setups where the vacuum
  configuration of \eq{eq:desired_vacuum} is the global minimum of the scalar
  potential whereas in the red regions, the charge-breaking vacuum
  configuration consists of a deeper minimum. 
 The blue lines separate the cases of a deeper DSB minimum and a deeper CB minimum
 in the context of the simplified model discussed above.
  The parameter space above 
  the black line features a tachyonic sneutrino at the tree level and scenarios in the grey 
  regions consequently violate $R$-parity while conserving (lighter grey) 
  or breaking (darker grey) electric charge.
  }
\label{fig:vrf_planes_vevacious}
\end{figure}

In \fig{fig:vrf_planes_vevacious}, we depict in the $(v_R,y_4^L)$ plane the
regions of the parameter space where the desired vacuum configuration
corresponds to the global minimum of the scalar potential (green) and the ones
where the charge-breaking configuration is preferred (red). We have found that
regions relevant for LHC physics (\textit{i.e.}, regions where the mass of the
doubly-charged Higgs is above the current limits) are not only exhibiting
a viable local minimum but also a global 
viable one. Moreover, as for the $y_4^L$-dependent
loop corrections to the \hppmm mass, larger the $y^L_4$ and $v_R$ values are,
the less favoured is the benchmark scenario under consideration.
Furthermore, as illustrated in the right panel of the figure,
a larger value of $\lambda_R \vs $ has the virtue of pulling the desired
vacuum configuration into a deeper minimum. However, as already above-mentioned,
this increases the possibility
of a tachyonic right sneutrino due to large negative
$F$-term contributions proportional to the $\lambda_R \vs$ product.
This happens at the tree level in the parameter space above the black line. In the region indicated by light grey shading we consequently find $R$-parity violating global minima, \textit{i.e.} $\langle \tilde \nu^c \rangle \neq 0$, which still conserve electric charge but feature too large electroweak vevs of the order of 
one TeV. 
In the dark grey area we find minima which break both, $R$-parity and electric charge. We stress that this $R$-parity violation is solely
due to the negative $F$-term contribution which drives a sneutrino tachyonic and could be evaded by a larger soft slepton mass. It is therefore of completely different origin than 
the $R$-parity violation suggested in early studies of this kind of models \cite{Kuchimanchi:1993jg,Huitu:1994zm}.
Finally, considering more than one right-handed neutrino leads to a
reduction of both the favoured region and the mass of the doubly-charged Higgs boson.
The blue lines in \fig{fig:vrf_planes_vevacious} show the separation between the red and the green area
as evaluated from the simplified model discussed in the beginning of this section. This simplified description can thus safely be used
to get rough information on which of the two minima is the deeper one. It does however not cover
the case of $R$-parity violating global minima as one would need to extend the simplified setup with vevs for the sneutrino fields.

Scenarios for which the global minimum is charge-breaking could nevertheless
be viable in cases where the tunneling time from the local $SU(2)_L\times U(1)_Y$
minimum to the global one is sufficiently large. 
The decay of a vacuum state into a deeper minimum can be described by the phase transition in which
bubbles of true vacuum nucleate out of a false vacuum state. For a successful phase transition, the
bubble has to be of a critical size and can, at zero temperature, be found by minimizing the Euclidean action
\cite{Coleman:1977py}
\begin{align}
S_E = \int {\rm d}^4x \Big(\frac{1}{2} (\partial_\mu \vec \phi)
(\partial^\mu \vec \phi) + V(\vec \phi) \Big)\,,
\end{align}
where $\vec \phi$ is a vector of all scalar fields and the coordinate $x^0$ corresponds to the imaginary time
coordinate $\tau = it$. Equivalently, the equations of motion have to be solved so that $\delta S_E=0$, which
eventually determines the optimal tunneling path in field space. For a first estimation of $S_E$,
the straight tunneling path between two vacua can be used, which corresponds to a reduction of the problem
to a one-dimensional problem. The latter can be solved numerically to arbitrary precision by the so-called
overshoot/undershoot method
(see, \textit{e.g.}, \REF{Wainwright:2011kj}). The full path deformation in all field dimensions is computationally 
extremely expensive and is hardly applicable to the dimensionality of the model studied in this work.
The decay rate of the false vacuum per unit volume $\mathcal V$ is then given by \cite{Coleman:1977py}
\begin{align}
\Gamma/\mathcal V = A e^{-S_E} \ .
\end{align}
In this expression, $A$ is a quantity whose dimension is in the fourth power of energy $E^4$ and
is related to the eigenvalues of a functional determinant~\cite{Callan:1977pt}. This determinant
is in practice usually estimated on dimensional 
grounds~\cite{Wainwright:2011kj}, $A \sim M^4$ with $M$ being the typical mass scale of the model.

A first estimate using the
direct tunneling path at zero temperature shows that all scenarios included in
the excluded region of the figure (red)
are metastable which is due to the large separation in field space of the different vacua.

It is however well known from the
MSSM~\cite{Camargo-Molina:2013sta,Camargo-Molina:2014pwa} that allowing for additional vevs and/or
including thermal effects can imply that metastable vacua are in fact unstable.
A thorough investigation of this effect lies beyond the scope of this paper.

Combining the results of the right panel of \fig{fig:vrf_planes_vevacious} with
the LHC constraint of \mbox{$v_R \gsim 4.9$~TeV} (see \SEC{sec:LHCbounds})
implies that all but a small strip of the parameter space turns out to be excluded
if one assumes that the right-handed neutrino is of a third generation nature. In
contrast, scenarios featuring a right-handed neutrino of one of the first two
generations turn out to be excluded.
In this way, we have demonstrated how accounting both for theoretical and
experimental constraints has allowed us to almost exclude all possible
setups in the left-right supersymmetric models under consideration.


\section{Conclusion}
\label{sec:conclusions}

We have studied a specific class of
supersymmetric models exhibiting a left-right gauge symmetry
and investigated to which extent experimental and theoretical constraints
restrict the viable regions of the parameter space.
A particular property of this class of models that feature Higgs fields lying
in the triplet representation of the $SU(2)_R$ group is that one of the
corresponding doubly-charged Higgs bosons gets tachyonic when the
scalar potential is minimized at the tree level. Even though it was known
that loop corrections taking into account the Yukawa couplings
of the right-handed neutrinos to the $SU(2)_R$ Higgs bosons were
modifying these conclusions, a complete \oneloop calculation
was still missing. We have filled this gap and shown that in a large
part of the parameter space, the complete one-loop calculation is necessary for
reliable predictions.

We have then studied to which extent the phenomenologically viable
regions of the parameter space (where the doubly-charged Higgs boson is
non-tachyonic), are constrained by
experimental data and theoretical considerations concerning the global minimum of the 
one-loop-corrected scalar potential. We have found that the latter favours lower values of
$v_R$ whereas the LHC gives tight lower bounds on this vev, in particular by 
exploring the $W_R$ decays into jets. Moreover, the lightest
doubly-charged Higgs boson is constrained to be well below 1~TeV for Yukawa couplings
imposed to lie in the perturbative regime. The left-right supersymmetric setup
that we have investigated is therefore close to exclusion, in particular if the
future LHC searches for a $W_R$ and a doubly-charged Higgs boson do not exhibit any signal
within the next few years.

\acknowledgments{
We thank Florian Staub and Ben O'Leary for useful discussions on \texttt{SARAH} and \texttt{Vevacious}
as well as
Ayon Patra for clarifications on their paper and Adam Alloul for discussions
in the early days of this project.
M.E.K and W.P. are supported by the DFG research training group
GRK1147 and by the DFG project no.\ PO-1337/3-1, L.B. and B.F. by
the French ANR 12 JS05 002 01 BATS@LHC and by the Theory-LHC-France
initiative of the CNRS/IN2P3 and Inphynity challenge of the CNRS/INP.
W.P. also thanks the CERN theory group, where part of this work has been carried out, for hospitality.

}


\providecommand{\href}[2]{#2}\begingroup\raggedright\endgroup

\end{document}